\documentclass[12pt,a4paper]{article}
\usepackage{a4}
\usepackage{amsmath}
\usepackage{amsfonts}
\usepackage{amssymb}

\usepackage{dsfont}
\usepackage{graphicx}
\usepackage{subfigure}

\allowdisplaybreaks[2]
\numberwithin{equation}{section}

\newcommand{\eq}[1]{\begin{equation}
                     \begin{split} #1 \end{split}
                     \end{equation}}

\newcommand{\bom}[1]{\fboxsep2mm\fbox{
           $ \displaystyle{ #1} $}}
           
\newcommand{\F}{\mathcal{F}}
\newcommand{\R}{\mathcal{R}}

\newcommand{\C}{\mathcal{C}}
\newcommand{\D}{\mathcal{D}}

\newcommand{\A}{\hat{\mathcal{A}}}
\newcommand{\ch}{\mbox{ch}}

\newcommand{\ov}[1]{\overline{#1}}

\begin{document}


\vspace*{-1.5cm}
\begin{flushright}
  {\small
  MPP-2008-145 
  }
\end{flushright}

\vspace{1.75cm}
\begin{center}
  {\LARGE
  The Generalized Green--Schwarz Mechanism \\
  for Type IIB Orientifolds with \\
  D3- and D7-Branes \\
  }
\end{center}

\vspace{0.5cm}
\begin{center}
  Erik Plauschinn 
\end{center}

\vspace{0.1cm}
\begin{center}
  \emph{
  Max-Planck-Institut f\"ur Physik \\ 
  F\"ohringer Ring 6, \\
  80805 M\"unchen, Germany \\
  } 
\end{center}

\vspace{0cm}
\begin{center}
  \tt{
  plausch\,@\,mppmu.mpg.de \\
  }
\end{center}

\vspace{1.5cm}
\begin{abstract}
\noindent In this paper, we work out in detail the tadpole cancellation conditions as well as the generalized Green--Schwarz mechanism for type IIB orientifold compactifications on smooth Calabi-Yau three-folds with D$3$- and D$7$-branes. We find that not only the D$3$- and D$7$-tadpole conditions have to be satisfied, but in general also the vanishing of the induced D$5$-brane charges leads to a non-trivial constraint. In fact, for the case $h^{1,1}_-\neq0$ the latter condition is important for the cancellation of chiral anomalies.  
We also extend our analysis by including D$9$- as well as D$5$-branes and  determine the rules for computing the chiral spectrum of the combined system.
\end{abstract}

\thispagestyle{empty}
\clearpage
\tableofcontents


\section{Introduction}

During the last years, the understanding of the open sector of type II string theories has grown to a mature state. 
In particular, on the type IIA side, where intersecting D$6$-branes allow for a geometric interpretation of the underlying structure, a set of rules for studying the low energy effective theory has been established \cite{Blumenhagen:2005mu} including consistency conditions such as the tadpole cancellation conditions and formulas for computing the chiral spectrum. Furthermore, it is known how to deal with anomalies via the generalized Green--Schwarz mechanism \cite{Green:1984sg,Sagnotti:1992qw,Aldazabal:1998mr,Ibanez:1998qp,Aldazabal:1999nu,Ibanez:1999pw,Scrucca:1999zh,Aldazabal:2000dg} and by imposing K-theory constraints \cite{Witten:1982fp,Uranga:2000xp}.
Within this framework, mostly on toroidal orbifolds, a huge number of models with various properties has been constructed.

However, orientifolds of type IIB string theory with D$9$- and D$5$-branes have been studied for an even longer time and are equally well-understood. Here, not intersecting branes but branes endowed with vector bundles are the objects of interest and,  similarly as on the IIA side, 
model building rules have been established (see for instance  \cite{Blumenhagen:2005zh} for the case of smooth Calabi-Yau compactifications) and a large number of models has been constructed.
Furthermore, T-duality allows to connect constructions on the type IIB and the type IIA side which in the past has helped to gain a better understanding for both descriptions. For a recent summary on the connection between type IIA models with D$6$-branes and type IIB models with D$9$-/D$5$-branes, including some \nolinebreak generalizations, see \cite{Bachas:2008jv}.

\bigskip
But type IIB string theory also allows for orientifold projections leading to configurations with D$3$- and D$7$-branes. Via T-duality, one naturally expects the open string sector to have the same features as the other two constructions which have been worked out for instance in \cite{Jockers:2004yj,Jockers:2005zy}. To our knowledge, however, some ingredients still require further study. In particular, although toroidal models are understood very well from a Conformal Field Theory point of view,
for a smooth compactification manifold the generalized Green--Schwarz mechanism has not been checked to work and also the tadpole cancellation conditions have not been derived in full detail.\footnote{The schematic form of the tadpole cancellation conditions for type IIB orientifolds with D$3$- and D$7$-branes from a geometric point of view has recently appeared in  \cite{Collinucci:2008pf}, however, here we study these conditions in detail.}

Let us emphasize this point: in this work, we focus solely on orientifold compactifications of string theory on {\em smooth} Calabi-Yau three-folds and formulate the effective theory in terms of topological quantities such as cycles and Chern characters. On the other hand, toroidal type IIB orientifolds generically contain singularities which are not suited for a geometric description but allow for a CFT formulation. For such configurations, the generalized Green--Schwarsz mechanism and the tadpole cancellation conditions are very well understood from a Conformal Field Theory point of view. Some of the references in this context  are \cite{Pradisi:1988xd,Gimon:1996rq,Angelantonj:1996uy,Kakushadze:1997ku,Aldazabal:1998mr,Ibanez:1998qp,Aldazabal:1999nu,Ibanez:1999pw,Scrucca:1999zh,Aldazabal:2000dg,Bianchi:2000de}.

\bigskip
As we have illustrated, from a phenomenological and geometrical point of view, the open string sector on smooth Calabi-Yau orientifolds is best understood on the type IIA side and on the type IIB side with D$9$- and D$5$-branes. The closed sector on the other hand is well-understood for type IIB orientifolds with D$3$- and D$7$-branes where the KKLT  \cite{Kachru:2003aw} and the Large Volume Scenarios \cite{Balasubramanian:2005zx,Conlon:2005ki} allow for a controlled study of closed string moduli stabilization. (For a recent discussion on the Large Volume Scenarios see \cite{Conlon:2008wa}.) 
However, as has been emphasized in \cite{Blumenhagen:2007sm}, moduli stabilization in the closed sector depends on the structure of the open sector and so it is necessary to understand also the latter for D$3$- and D$7$-branes in more detail.

Furthermore, F-theory \cite{Vafa:1996xn} provides a description of type IIB string theory with D$3$- and D$7$-branes beyond the perturbative level which has recently become of interest for phenomenology \cite{Beasley:2008dc,Buchbinder:2008at,Beasley:2008kw,Heckman:2008es,Marsano:2008jq,Donagi:2008kj,Marsano:2008py,Heckman:2008qt,Wijnholt:2008db,Font:2008id,Heckman:2008qa,Blumenhagen:2008zz} (see also \cite{Braun:2008pz}). Although the constructions in this context concentrate mostly on local models, at some point these have to be embedded into a compact manifold implying for instance that the tadpole cancellation conditions have to be satisfied.

\bigskip
The outline and the results of this work are summarized as follows.
In section \ref{sec_tadpole}, we derive the tadpole cancellation conditions for type IIB string theory compactified on orientifolds of smooth Calabi-Yau three-folds with D$3$- and D$7$-branes.
In addition to the well-known D$3$- and D$7$-brane tadpoles, we also work out the cancellation conditions for induced D$5$-brane charges. 
In section \ref{sec_chiral_anom}, we briefly summarize the expressions for the chiral anomalies in the present context, and in section \ref{sec_gsm}, we show that the generalized Green--Schwarz mechanism indeed cancels the anomalies using the tadpole cancellation conditions. In particular, we emphasize that in general the induced D$5$-brane charge conditions have to be employed. In section \ref{sec_generalization}, we generalize our analysis by including D$9$- as well as D$5$-branes for which we work out the tadpole cancellation conditions and the formulas for computing the chiral spectrum. 
In section \ref{sec_summary}, we finish with some conclusions.


\section{Tadpole Cancellation Conditions}
\label{sec_tadpole}


\subsection{Setup and Notation}

Before deriving the tadpole cancellation conditions, let us first make clear the setup we are working in and recall some results needed for the following. 


\subsubsection*{Orientifold Compactification}

We consider type IIB string theory compactifications from a ten-dimensional space-time to four dimensions on a compact Calabi-Yau three-fold $\mathcal{X}$
\eq{
  \label{compactf}
  \mathbb{R}^{9,1} \to \mathbb{R}^{3,1} \times \mathcal{X} \;.
}
In order to introduce D-branes and break supersymmetry to $\mathcal{N}=1$ in four dimensions, we also perform an orientifold projection $\Omega (-1)^{F_L} \sigma$ where $\Omega$ is the world-sheet parity operator, $F_L$ is the left-moving fermion number and $\sigma$ is a holomorphic involution on $\mathcal{X}$. The action of $\sigma$ on the K\"ahler form $J$  and the holomorphic three-form $\Omega_3$ of $\mathcal{X}$ is chosen to be
\eq{
  \label{orient_choice}
  \sigma^*J=+J\;,\hspace{40pt}
  \sigma^*\Omega_3=-\Omega_3\;,
}
allowing for O$3$- and O$7$-planes. 
The action of $\Omega (-1)^{F_L}$ on the metric $g$, the dilaton $\phi$, the Neveu Schwarz-Neveu Schwarz (NS-NS) two-form $B_2$, the gauge invariant open string field strength $\F$ and the  Ramond-Ramond (R-R) $p$-form potentials $C_p$ is determined to be of the following form \cite{Acharya:2002ag,Brunner:2003zm,Grimm:2004uq}
\eq{
  \label{orient_signs}
  \renewcommand{\arraystretch}{1.25}
  \begin{array}{lclclcl}
  \arraycolsep1.5pt
  \displaystyle \Omega\, (-1)^{F_L}\: g &=& + \:g \;,  & \qquad &
  \displaystyle \Omega\, (-1)^{F_L}\: \mathcal{F} &=& -\:\mathcal{F} \;, \\
  \displaystyle \Omega\, (-1)^{F_L}\: \phi &=& + \:\phi \;,  & \qquad &
  \displaystyle \Omega\, (-1)^{F_L}\: C_p &=& (-1)^{\frac{p}{2}} 
    \:C_p \;, \\
  \displaystyle \Omega\, (-1)^{F_L}\: B_2 &=& - \:B_2 \;.  & \qquad &
  \end{array}
}
Note that we are going to work with the democratic formulation of type IIB supergravity \cite{Bergshoeff:2001pv} so that the R-R $p$-form potentials $C_p$ appear for $p=0,2,4,6,8(,10)$.


\subsubsection*{(Co-)Homology}

The holomorphic involution $\sigma$ introduced above gives rise to a splitting of the cohomology groups $H^{p,q}(\mathcal{X},\mathbb{Z})$ into 
the even and odd eigenspaces of $\sigma^*$ (here we mainly follow \cite{Grimm:2004uq}) 
\eq{
  \label{cohom_split}
  H^{p,q} = 
  H^{p,q}_+\oplus
  H_-^{p,q} \;.
}
The dimensions of these spaces are denoted by $h^{p,q}_{\pm}$ for which the following relations can be determined \cite{Grimm:2004uq}
\eq{
  \label{cohom_split_2}
  \arraycolsep2pt
  \renewcommand{\arraystretch}{1.25}
  \begin{array}{lcllclcllcl}
  h^{1,1}_{\pm} & = & h^{2,2}_{\pm} \;, \qquad\;\;
    h^{3,0}_{+} & = & h^{0,3}_{+} & = & 0 \;, \qquad\;\;
      h^{0,0}_{+} & = & h^{3,3}_{+} & = & 1 \;, \\
  h^{2,1}_{\pm} & = & h^{1,2}_{\pm} \;, \qquad\;\;
    h^{3,0}_{-} & = & h^{0,3}_{-} & = & 1 \;, \qquad\;\;
      h^{0,0}_{-} & = & h^{3,3}_{-} & = & 0 \;.      
  \end{array}
}

Next, let us introduce some notation for the third (co-)homology group of $\mathcal{X}$. In particular, we denote a basis of three-cycles on $\mathcal{X}$ by $\{\alpha_i,\beta^j\}\in H_3(\mathcal{X},\mathbb{Z})$ where $i,j=0,\ldots,h^{2,1}$. This basis can be chosen in such a way, that the Poincar\'{e} duals $\{[\alpha_i],[\beta^j]\}$ satisfy
\eq{
  \label{flux_quant_3}
  \int_{\mathcal{X}}\,[\alpha_i]\wedge [\beta^j]=l_s^6\,\delta_i^j
  \;,\hspace{40pt}
  \int_{\mathcal{X}}\,[\alpha_i]\wedge [\alpha_j]
  =\int_{\mathcal{X}}\,[\beta^i]\wedge [\beta^j]=0 \;,
}
where $l_s$ denotes the string length. Note that, as indicated in \eqref{cohom_split}, these relations decompose into the even and odd eigenspaces of $\sigma^*$, which means that the only non-trivial relations are
\eq{
  \int_{\mathcal{X}}\,[\alpha^{\pm}_i]\wedge 
  [\beta^{\pm j}]=l_s^6\,\delta_i^j 
  \hspace{40pt}\mbox{with}\hspace{40pt}
  \left\{
  \begin{array}{lcc}
  i,j=1,\ldots,h^{2,1}_+ & \mbox{for} & + \;, \\
  i,j=0,\ldots,h^{2,1}_- & \mbox{for} & - \;, 
  \end{array}
  \right.
}    
where $\pm$ labels elements of the even respectively odd co(-homology) group.
To continue, we denote a basis of $(1,1)$- and $(2,2)$-forms on $\mathcal{X}$ as
\eq{
  \label{basis_11}
  \bigl\{ \omega_I\bigr\} \in H^{1,1}\bigl( \mathcal{X},\mathbb{Z} \bigr)
  \;,\hspace{40pt}
  \bigl\{ \sigma^I\bigr\} \in H^{2,2}\bigl( \mathcal{X},\mathbb{Z} \bigr)
  \;.
}
These two basis will be chosen such that
\eq{
  \int_{\mathcal{X}} \omega_I\wedge \sigma^J = l_s^6\,\delta_I^J 
   \;,
}  
where the index $I$ takes values $I=1,\ldots,h^{1,1}(\mathcal{X})$ and, similarly as above, this relation decomposes into the even and odd eigenspaces of $\sigma^*$.
Finally, we introduce a basis of four- and two-cycles  on $\mathcal{X}$
\eq{
  \label{basis_4_cycles}
  \bigl\{ \gamma_I \bigr\} \in H_4 \bigl( \mathcal{X},\mathbb{Z} \bigr)
  \;,\hspace{40pt}
  \bigl\{ \Sigma^I \bigr\} \in H_2 \bigl( \mathcal{X},\mathbb{Z} \bigr)
  \;,
}
in such a way that the Poincar\'e duals of $\gamma_I$ and $\Sigma^I$ are $[\gamma_I]=\frac{1}{l_s^2}\:\omega_I$ respectively $[\Sigma^I]=\frac{1}{l_s^4}\:\sigma^I$. Concretely, this means  that
\eq{
  \int_{\gamma_I} \sigma^J 
  = l_s^4 \: \delta^J_I 
  \;,\hspace{40pt}
  \int_{\Sigma^I} \omega_J 
  = l_s^2 \: \delta_J^I \;.
}


\subsubsection*{Background Fluxes}

Let us now consider the closed string sector in some more detail. In particular, we are allowed to turn on supersymmetric background fluxes in $\mathcal{X}$, i.e. we can have non-vanishing VEVs for \cite{Giddings:2001yu}
\eq{
  F_3 = d C_2
  \qquad\mbox{and}\qquad
  H_3 = d B_2 \;.
}  
Because of the Dirac quantization condition, such fluxes are quantized. Furthermore,  since we perform an orientifold projection $\Omega(-1)^{F_L}\sigma$, there are some subtleties due to the involution $\sigma$ on $\mathcal{X}$ \cite{Kachru:2002he,Frey:2002hf}. Although these issues can be dealt with, here we stay on firm grounds and impose the following quantization conditions
\eq{
  \label{flux_quant}
  \arraycolsep1.5pt
  \begin{array}{lclclcl}
  \displaystyle 
  \frac{1}{l_s^2} \int_{\alpha^-_{i}} F_3 &=& 2f_i\;\in 2\mathbb{Z} \;,
  &\hspace{40pt}&
  \displaystyle 
  \frac{1}{l_s^2} \int_{\beta^{-j}} F_3 &=& 2f^j\;\in 2\mathbb{Z} \;, 
  \\[5mm]
  \displaystyle 
  \frac{1}{l_s^2} \int_{\alpha^-_i} H_3 &=& 2h_i\;\in 2\mathbb{Z} \;,
  &\hspace{40pt}&
  \displaystyle 
  \frac{1}{l_s^2} \int_{\beta^{-j}} H_3 &=& 2h^j\;\in 2\mathbb{Z} \;,
  \end{array}
}
with $i,j=0,\ldots,h^{2,1}_-$. Note that because $F_3$ and $H_3$ are odd under $\Omega(-1)^{F_L}$, we only turn on flux through cycles $\{\alpha^-_i,\beta^{-j}\}$ odd under the orientifold projection. 
Using then \eqref{flux_quant} and \eqref{flux_quant_3}, we can express $F_3$ and $H_3$ in the following way 
\eq{
  \label{flux_quant_2}
  F_3=\frac{2}{l_s}\Bigl( f^i\,\bigl[\alpha^-_i\bigr]-f_j\,
  \bigl[\beta^{-j}\bigr]\Bigr)
  \;,\hspace{40pt}
  H_3=\frac{2}{l_s}\Bigl( h^i\,\bigl[\alpha^-_i\bigr]-h_j\,
  \bigl[\beta^{-j}\bigr]\Bigr)\;.
}


\subsubsection*{D-Branes and Gauge Fluxes}

After having discussed fluxes in the closed sector, we now turn to the open sector. The fixed loci of the involution $\sigma$ on $\mathcal{X}$ are called orientifold planes and for the choice \eqref{orient_choice}, these are O$3$- and O$7$-planes usually carrying negative R-R and NS-NS charges. Therefore, as we will see below, we have to introduce a combination of D$3$-branes and background flux as well as D$7$-branes wrapping 
\eq{
  \mbox{holomorphic divisors}\quad \Gamma_{{\rm D}7}\quad \mbox{in}
  \quad \mathcal{X} \;. 
}

\bigskip
It is furthermore possible to turn on gauge flux $\ov F$ on the D$7$-branes which, in order to preserve supersymmetry, has to obey the constraints \cite{Dasgupta:1999ss,Marino:1999af,Jockers:2005zy}
\eq{
  \ov{F}^{(2,0)}=\ov{F}^{(0,2)}=0\;,
  \hspace{50pt}
  \bigl( J\wedge\ov F\,\bigr) \Bigl\lvert_{\Gamma_{{\rm D}7}}=0 \;.
}
Moreover, to preserve four-dimensional Lorentz invariance, we consider gauge flux  $\ov F$ only in the internal space $\mathcal{X}$ and so we make the following ansatz for the total open string field strength $\mathbf{F}$
\eq{
  \label{flux_split}
  \mathbf{F} = F + \ov  F
}
with $F$ denoting the field strength of the gauge field in $\mathbb{R}^{3,1}$ while $\ov F$ stands for the flux components in $\mathcal{X}$. However, $\mathbf{F}$ is not gauge invariant and so we define 
\eq{
  \label{def_f}
  \F = -i\,\bigl(\, l_s^2\,\mathbf{F} + 2\pi \varphi^* B_2\,\mathds{1}
  \,\bigr) \;,
}
which we call the gauge invariant open string field strength.
In \eqref{def_f}, $l_s$ denotes again the string length, $B_2$ is the NS-NS two-form and $\varphi^*$ is the pull-back from $\mathcal{X}$ to the holomorphic divisor $\Gamma_{{\rm D}7}$ the D$7$-brane is wrapping. Note that we will also employ the notation 
\eq{
  \label{def_f_bar}
  \ov \F \qquad\ldots\qquad\mbox{components of}\quad \F\quad 
  \mbox{in}\quad \mathcal{X} \;.
}

To conclude our discussion of the open string gauge fluxes, let us split the NS-NS two-form $B_2$ on $\mathcal{X}$ into parts which are even respectively odd under the holomorphic involution $\sigma$
\eq{
  \label{notation_10}
  B^{(6)}_2 = B_2^+ + B_2^-
  \;,\hspace{60pt}
  \sigma^*  B_2^{\pm} = \pm \,B_2^{\pm} \;,
}
where \textsuperscript{(6)} denotes the components of $B_2$ in $\mathcal{X}$.
Due to the action of $\Omega(-1)^{F_L}$, $B_2^+$ has to take discrete values which is important for the correct quantization of 
\eq{
  \label{notation_11}
  \ov \F^+ = -i\,\bigl(\, l_s^2\,\ov F +2\pi \, 
  \varphi^*B^+_2\,\mathds{1}\bigr) 
}
(see for instance \cite{Bachas:2008jv} and references therein).
The components $B_2^-$ on the other hand, are part of the moduli $G^{I_-}$ for $I_-=1,\ldots,h^{1,1}_-$ \cite{Grimm:2004uq,Jockers:2004yj} (see also \cite{Hristov:2008if}) and take continuous values.

\medskip
Finally, we denote the total curvature two-form of the tangent bundle of $\mathbb{R}^{3,1}\times\mathcal{X}$ by $\mathbf{R}$ which splits into $R$ and $\ov R$  defined on $\mathbb{R}^{3,1}$ respectively $\mathcal{X}$. For dimensional reasons, we then define
\eq{
  \label{def_r}
  \R = l_s^2\,\mathbf{R} = l_s^2\,\bigl( R + \ov R\,\bigr) \;.
}


\subsection{Effective Actions}

As we have already mentioned, in order to study D$3$- and D$7$-branes, it is useful to work with the democratic formulation of type IIB supergravity in ten dimensions. The  bosonic part of this (pseudo-)action reads \cite{Bergshoeff:2001pv}
\eq{
  \label{action_iib}
  &\mathcal{S}_{\rm IIB} = \frac{1}{2\kappa_{10}^2} 
  \int \biggl[\, e^{-2\phi}\biggl( \mathsf{R}\star1 
  + 4\,d\phi\wedge\star d\phi 
  -\frac{1}{2}\, H_3\wedge\star H_3\biggr)  \\
  &\hspace{208pt}
  -\frac{1}{4} \sum_{p=1,3,5,7,9} \widetilde F_p\wedge \star 
  \widetilde F_p \,\biggr] \;,
}
where $(2\kappa_{10}^2)^{-1}\!=2\pi\,l_s^{-8}$, the star $\star$ stands for the Hodge-$\star$-operator and $\mathsf{R}$ denotes the  curvature scalar. The generalized field strengths $\widetilde F_p$ together with their duality relations take the following form
\eq{
  \label{duality}
  \widetilde F_p = d\,C_{p-1}-H_3\wedge C_{p-3} 
  \;,\hspace{60pt}
  \widetilde F_p = (-1)^{\frac{p+3}{2}}\star 
  \widetilde F_{10-p}\;. 
}  
Later, we will focus on the equation of motion for the Ramond-Ramond (R-R)  fields $C_8$, $C_6$ and $C_4$ and so we calculate the variation of \eqref{action_iib} with respect to these fields
\eq{
  \label{variation_closed}
  \delta_{C_4}\mathcal{S}_{\rm IIB} &= 
  \frac{1}{4\kappa_{10}^2} \int \delta C_4 \wedge\Bigl( +d \widetilde F_5 
  - H_3\wedge \widetilde F_3 \Bigr)\;, \\
  \delta_{C_6}\mathcal{S}_{\rm IIB} &=
  \frac{1}{4\kappa_{10}^2} \int \delta C_6 \wedge\Bigl( -d \widetilde F_3 
  + H_3\wedge \widetilde F_1 \Bigr)\;, \\
  \delta_{C_8}\mathcal{S}_{\rm IIB} &=
  \frac{1}{4\kappa_{10}^2} \int \delta C_8 \wedge\Bigl( +d \widetilde F_1 
  \Bigr)\;.
}
Since we only turn on fluxes $F_3$ and $H_3$, the term $H_3\wedge \widetilde F_1$ vanishes. For the first line in \eqref{variation_closed}, we employ \eqref{flux_quant_2} and \eqref{flux_quant_3} to calculate
\eq{
   \label{flux_number}
   \frac{1}{l_s^4} \int_{\mathcal{X}} H_3\wedge \widetilde F_3 =
   \frac{1}{l_s^4} \int_{\mathcal{X}} H_3\wedge F_3 =
   4\,\bigl(h_i\,f^i- h^i\, f_i\bigr) =
   -4N_{\rm flux} \;\in 4\mathbb{Z}
   \;,
}
where the minus sign has been chosen for later convenience.

\bigskip  
After having discussed the closed sector, we now turn to the open sector for which the Chern-Simons action of the D$p$-branes and O$p$-planes read 
\cite{Douglas:1995bn,Green:1996dd,Cheung:1997az,Morales:1998ux,Stefanski:1998yx,Scrucca:1999uz} (see also \cite{Blumenhagen:2006ci})
\eq{
  \label{action_cs}
  \mathcal{S}^{\rm CS}_{{\rm D}p} &= -\mu_p \int_{{\rm D}p} 
    \mbox{ch}\left( \F\right) \wedge
    \sqrt{ \frac{\A(\R_T)}{\A(\R_N)}} \wedge
    \bigoplus_q \varphi^*C_q \;, \\
  \mathcal{S}^{\rm CS}_{{\rm O}p} &= -\,Q_p\, \mu_p \int_{{\rm O}p} 
    \sqrt{ \frac{\mathcal{L}(\R_T/4)}{\mathcal{L}(\R_N/4)}} \wedge
    \bigoplus_q \varphi^* C_q \;,
}
Here, $\varphi^*$ denotes again the pull-back from $\mathcal{X}$ to the manifold the D-brane respectively O-plane is wrapping and $\R_T$, $\R_N$ stand for the restrictions of $\R$ to the tangent and normal bundle of this manifold. Furthermore, in the present case we have \footnote{The signs $\kappa_p=\pm1$ in \eqref{tension} have already appeared in \cite{Blumenhagen:2005zh}, where they were crucial in order to obtain the correct matching between the tadpole cancellation conditions of type IIB orientifolds with D$9$-/D$5$-branes and the anomaly cancellation condition of the heterotic string. Similarly, here the signs are important to match the D$3$-brane tadpole cancellation condition with F-theory.}
\eq{
  \label{tension}
  \mu_p=\frac{2\pi}{l_s^{p+1}}\:\kappa_p
  \hspace{40pt}\mbox{with}\hspace{40pt}
  \begin{array}{c}
    \kappa_7=+1\;, \\
    \kappa_3=-1 \;,
    \end{array}
}
and the sums in \eqref{action_cs} run over $q=0,2,4,6,8,10$. The charge of the O$p$-planes is given by $Q_p=-2^{p-4}$.

The definition of the Chern character, the $\A$ genus and Hirzebruch polynomial $\mathcal{L}$ can be found in appendix \ref{app_cs_action} together with the calculation leading to the following expressions 
\eq{
  \label{top_exp_1}
  \arraycolsep1.5pt
  \renewcommand{\arraystretch}{2.0}  
  \begin{array}{llclcl}
  {\rm D}3:\quad & 
    \sqrt{ \frac{\A\left(\R_T\right)}{\A\left(\R_N\right)}}  &
    = & 
     \left( 1 + \frac{1}{96} \left(\frac{l_s^2}{2\pi}\right)^2
    \!\mbox{tr}\,\bigl( R^2\bigr)
    +\ldots\hspace{6pt}\right),\!\!\!
    \\
  {\rm D}7:\quad &
    \sqrt{ \frac{\A\left(\R_T\right)}{\A\left(\R_N\right)}}  &
    = & 
    \left( 1 + \frac{1}{96} \left(\frac{l_s^2}{2\pi}\right)^2
    \!\mbox{tr}\,\bigl( R^2\bigr)
    +\ldots\hspace{6pt}\right)
    &\wedge &
    \biggl( 1 + \frac{l_s^4}{24}\: c_2\bigl( \Gamma_{{\rm D}7}\bigr) + 
    \ldots\biggr),
    \\
  {\rm O}3:\quad & 
    \sqrt{ \frac{\mathcal{L}\left(\R_T/4\right)}{\mathcal{L}
    \left(\R_N/4\right)}}  &
    = & 
     \left( 1 - \frac{1}{192} \left(\frac{l_s^2}{2\pi}\right)^2
    \!\mbox{tr}\,\bigl( R^2\bigr)
    +\ldots\right),\!\!\!
    \\
  {\rm O}7:\quad &
    \sqrt{ \frac{\mathcal{L}\left(\R_T/4\right)}{\mathcal{L}
    \left(\R_N/4\right)}}  &
    = & 
    \left( 1 - \frac{1}{192} \left(\frac{l_s^2}{2\pi}\right)^2
    \!\mbox{tr}\,\bigl( R^2\bigr)
    +\ldots\right) &
    \wedge &
    \biggl( 1 - \frac{l_s^4}{48}\: c_2\bigl( \Gamma_{{\rm O}7}\bigr) + 
    \ldots\biggr) .
  \end{array}
}
Note that $R$ is defined on $\mathbb{R}^{3,1}$ and that the four-form $c_2$ is  defined on $\mathcal{X}$. Also, we have only shown the terms relevant for the integrals in the Chern-Simons actions. 
With the help of \eqref{top_exp_1}, we can now compute the variation of the actions \eqref{action_cs} with respect to $C_4$, $C_6$ and $C_8$. We find
\eq{
  \label{variation_4}
  \delta_{C_4} \mathcal{S}^{\rm CS}_{{\rm D}3}
    &= +\mu_3\int_{\mathbb{R}^{3,1}} \delta C_4 \, N_{{\rm D}3}\;,
    \\
   \delta_{C_4}  \mathcal{S}^{\rm CS}_{{\rm O}3}
    &= +\mu_3\int_{\mathbb{R}^{3,1}} \delta C_4\,\left(-\frac{1}{2}\right)
    \;,
    \\
   \delta_{C_4} \mathcal{S}^{\rm CS}_{{\rm D}7}
    &= -\mu_7\int_{\mathbb{R}^{3,1}}
    \delta C_4\wedge\int_{\Gamma_{{\rm D}7}}\left( \ch_2\bigl( \ov 
    \F\bigr) 
    +l_s^4\, N_{{\rm D}7} \,\frac{c_2\bigl( \Gamma_{{\rm D}7}\bigr) }{24} 
    \right)
    \;,
    \\
  \delta_{C_4} \mathcal{S}^{\rm CS}_{{\rm O}7}
    &= -\mu_7\int_{\mathbb{R}^{3,1}} 
    \delta C_4\wedge\int_{\Gamma_{{\rm O}7}} \:
    l_s^4\;\frac{c_2\bigl(\Gamma_{{\rm O}7}\bigr)}{6}
    \;,
}  
with $N_{{\rm D}3}=\ch_{0}\bigl(\F_{{\rm D}3}\bigr)$ and $N_{{\rm D}7}=\ch_{0}\bigl(\F_{{\rm D}7}\bigr)$ denoting the number of D$3$- respectively D$7$-branes on top of each other. 
Furthermore, $\Gamma$ is again the holomorphic four-cycle wrapped by the D$7$-branes and O$7$-planes in the compact space, and $\ov\F$ stands for the  part of $\F$ in $\mathcal{X}$.
In a similar way as above, we compute the variation of the Chern-Simons actions with respect to  $C_6$ as follows
\eq{
  \label{variation_6}  
  \delta_{C_6}\mathcal{S}^{\rm CS}_{{\rm D}7}
    = -\mu_7\int_{\mathbb{R}^{3,1}\times\Gamma_{{\rm D}7}} 
    \bigl(\varphi^*\delta  C_6\bigr)\: \wedge\ch_1\bigl( \ov\F\bigr) 
    \;,
    \hspace{40pt}
  \delta_{C_6} \mathcal{S}^{\rm CS}_{{\rm O}7}
    = 0
    \;,
}  
and the variation with respect to $C_8$ is found to be
\eq{
  \label{variation_8}
  \delta_{C_8} \mathcal{S}^{\rm CS}_{{\rm D}7}
    = -\mu_7\int_{\mathbb{R}^{3,1}\times\Gamma_{{\rm D}7}} 
   \bigl(\varphi^*\delta  C_8\bigr)\:N_{{\rm D}7}
    \,,
    \qquad
  \delta_{C_8} \mathcal{S}^{\rm CS}_{{\rm O}7}
    = -\mu_7 \int_{\mathbb{R}^{3,1}\times\Gamma_{{\rm O}7}} 
    \bigl(\varphi^*\delta  C_8\bigr)\: \bigl( - 8\bigr)
    \,.
}


\subsection{Tadpole Cancellation Conditions}

Combining the results from the previous subsection, we can now determine the tadpole cancellation conditions for type IIB orientifolds with D$3$- and D$7$-branes. However, because of the orientifold projection $\Omega (-1)^{F_L} \sigma$, we have to take into account the orientifold planes as well as the orientifold images of the D-branes. Denoting these images by a prime, the schematic form of the full action is 
\eq{
  \label{full_action}
  \mathcal{S} = 
  \frac{1}{2}\left( 
  2\:\mathcal{S}_{\rm IIB} 
  + \sum_{a,a'} \mathcal{S}^{\rm CS}_{{\rm D}7_a}
  + \sum_{i} \mathcal{S}^{\rm CS}_{{\rm O}7_i}
  + \sum_{b,b'} \mathcal{S}^{\rm CS}_{{\rm D}3_b}
  + \sum_{j} \mathcal{S}^{\rm CS}_{{\rm O}3_j}
  \right)
  \;.
}
In order to be more concrete later, using \eqref{orient_signs}, we determine the data for an orientifold image of a D-brane as follows
\eq{
  \label{oaction_manif}
  \arraycolsep2pt
  \renewcommand{\arraystretch}{1.5}
  \begin{array}{lclcl}
  \displaystyle \Gamma'_{{\rm D}p} &=& 
  \displaystyle \Omega \, (-1)^{F_L}\sigma\,\Gamma_{{\rm D}p}
  &=& \displaystyle  (-1)^{\frac{p+1}{2}}\: \sigma\,\Gamma_{{\rm D}p}
  \;, \\
  \displaystyle {\ov\F^+}' &=& 
  \displaystyle \Omega \, (-1)^{F_L}\sigma^*\,\ov\F^+ &=&
  \displaystyle -\, \sigma^*\ov\F^+\;,
  \end{array}
}
where $\Gamma_{{\rm D}7}$ is a holomorphic divisor in $\mathcal{X}$ while $\Gamma_{{\rm D}3}$ is a point in $\mathcal{X}$ corresponding to a D$3$-brane.


\subsubsection*{D$\mathbf 7$-Brane Tadpole Cancellation Condition}

The equations of motion for $C_8$ are obtained by setting to zero the variation of \eqref{full_action} with respect to $C_8$. Using \eqref{variation_closed} and \eqref{variation_8}, we  compute
\eq{
  0&= \frac{1}{2}\:
  \frac{2\pi}{l_s^8}\int_{\mathbb{R}^{3,1}\times\mathcal{X}} 
  \delta  C_8\wedge\Biggl[\,
   d\widetilde F_1
    - \frac{1}{l_s^2}\sum_{\,{\rm D}7_a,{\rm D}7_{a'}} N_{{\rm D}7_a} 
    \bigl[ \Gamma_{{\rm D}7_a} \bigr]
    -\frac{1}{l_s^2}\sum_{{\rm O}7_i} \bigl( -8 \bigr)
    \bigl[ \Gamma_{{\rm O}7_i} \bigr]
  \Biggr] ,
}
where $N_{{\rm D}7_a}$ is the total number of D$7$-branes with gauge flux $\ov F_a$ wrapping the four-cycle $\Gamma_{{\rm D}7_a}$, and 
$[\Gamma]$ stands for the Poincar\'{e} dual of the four-cycle $\Gamma$ in $\mathcal{X}$.  
Since the variations $\delta C_8$ are arbitrary and $d \widetilde F_1$ is exact, in cohomology the expression above can be written as
\eq{
  \label{tadpole_8}
  \bom{
  \sum_{{\rm D}7_a} N_{{\rm D}7_a}
  \Bigl(\, [\Gamma_{{\rm D}7_a}] +[\Gamma'_{{\rm D}7_a}]\,\Bigr) = 8 \sum_{{\rm O}7_i}\: 
  [\Gamma_{{\rm O}7_i}]\;,
  }
}
which is known as the D$7$-brane tadpole cancellation condition.


\subsubsection*{D$\mathbf 5$-Brane Tadpole Cancellation Condition}

Employing \eqref{variation_closed} and \eqref{variation_6} as well as
the basis of $(1,1)$-forms $\{\omega_I\}\in H^{1,1}( \mathcal{X},\mathbb{Z})$ introduced in equation \eqref{basis_11}, the equations of motion originating from $C_6$ are found to be of the following form
\eq{
  \label{tadpole_6_pre}
  0&= \sum_{{\rm D}7_a} \left(
    \int_{\Gamma_{{\rm D}7_a}} \ch_1\bigl(\ov\F_a\bigr) \wedge
    \omega_I
    +
    \int_{\Gamma'_{{\rm D}7_{a}}}\ch_1\bigl(\ov\F'_a\bigr) 
    \wedge \omega_I \right) \\
  &= \sum_{{\rm D}7_a} \int_{\mathcal{X}}\biggl(
    \ch_1\bigl(\varphi_*\ov\F_a\bigr) \wedge
    \left[ \Gamma_{{\rm D}7_a} \right]
    +
    \ch_1\bigl(\varphi_*\ov\F'_a\bigr) \wedge
    \left[ \Gamma'_{{\rm D}7_{a}} \right]  
    \biggr)\wedge \omega_I 
}
where the prime denotes again the $\Omega (-1)^{F_L} \sigma$ image and  $\varphi_*\ov\F$ is the push-forward of $\ov\F$ from the D$7$-brane to the Calabi-Yau manifold $\mathcal{X}$.

\bigskip
Note that \eqref{tadpole_6_pre} is not trivially vanishing which can be seen by utilizing  the relation 
\eq{
  \label{int_oim}
  \int_{\mathcal{X}} \sigma^* \omega_I\wedge
    \sigma^* \omega_J\wedge \sigma^* \omega_K
  = \int_{\mathcal{X}}  \omega_I\wedge
    \omega_J\wedge \omega_K \;.
}
In particular, recalling from \eqref{orient_signs} that $\F$ is odd under $\Omega(-1)^{F_L}$, we can rewrite \eqref{tadpole_6_pre} as
\eq{
  0 = \sum_{{\rm D}7_a} 
    \ch_1\bigl(\varphi_*\ov\F_a\bigr) \wedge
    \left[ \Gamma_{{\rm D}7_a} \right]
    \wedge \Bigl( \omega_I - \sigma^* \omega_I \Bigr) \;,
}    
which is a non-trivial constraint if $h^{1,1}_-\neq0$.
\pagebreak[3]

\bigskip
However, the D$5$-brane tadpole cancellation condition is not yet satisfying.\footnote{We thank the authors of \cite{Blumenhagen:2008zz}  for pointing out this issue to us. The discussion in this paragraph is based on work in \cite{Blumenhagen:2008zz}, which we present here in order to give a consistent derivation of the tadpole cancellation conditions.} In order to explain this point, let us recall from our discussion around equation \eqref{notation_10} that $\ov\F$ contains $B_2^-$ which takes continuous values. Since the tadpole cancellation conditions usually involve only discrete quantities, the dependence on $B_2^-$ should disappear. And indeed, using the definition of the Chern character \eqref{def_ch} as well as \eqref{notation_10} and \eqref{notation_11}, we compute
\eq{
   \label{ch_1_B2}
   \ch_1\bigl(\varphi_*\ov\F_a\bigr) 
   = \ch_1\Bigl(  \varphi_*\ov\F^+_a \Bigr) 
   + N_{{\rm D}7}\, B_2^-\;.
}
Employing then the D$7$-brane tadpole cancellation condition \eqref{tadpole_8}, we find for the $B_2^-$ terms in \eqref{tadpole_6_pre} that
\eq{
  \sum_{{\rm D}7_a} \int_{\mathcal{X}}\biggl(
    N_{{\rm D}7_a}\, B_2^- \wedge
    &\left[ \Gamma_{{\rm D}7_a} \right]
    +
    N_{{\rm D}7_a}\, B_2^- \wedge
    \left[ \Gamma'_{{\rm D}7_a} \right]  
    \biggr)\wedge \omega_I \\
  =& \:8 \,\sum_{{\rm O}7_i}\:\int_{\mathcal{X}} [\Gamma_{{\rm O}7_i}]
    \wedge B_2^- \wedge \omega_I
  = 8 \,\sum_{{\rm O}7_i}\:\int_{\Gamma_{{\rm O}7_i}} 
    \varphi^*B_2^- \wedge \omega_I \;. 
  \label{tadpole_5_10}
}
The final step is to observe that since the orientifold planes are pointwise invariant under the involution $\sigma$, there are no odd two-cycles on $\Gamma_{{\rm O}7}$, that is $H_{2-}(\Gamma_{{\rm O}7},\mathbb{Z})=0$. Because $B_2^-$ is in $H^{1,1}_-(\mathcal{X},\mathbb{Z})$, we see that in this case $\varphi^* B_2^-=0$ and so the integral \eqref{tadpole_5_10} vanishes.
The D$5$-brane tadpole cancellation condition therefore contains only discrete quantities and reads
\eq{
  \label{tadpole_6}
  \bom{
  0= \sum_{{\rm D}7_a} \biggl(
    \ch_1\Bigl(\varphi_*\ov\F^+_a\Bigr) \wedge
    \left[ \Gamma_{{\rm D}7_a} \right]
    +
    \ch_1\Bigl(\varphi_*{\ov\F^+}'_a\Bigr) \wedge
    \left[ \Gamma'_{{\rm D}7_a} \right]  
    \biggr)\wedge \omega_I \;.
  }
}


\subsubsection*{D$\mathbf 3$-Brane Tadpole Cancellation Condition}

Let us finally study the equation of motion for $C_4$ which is obtained by setting to zero the variation of \eqref{full_action} with respect to $C_4$. Employing \eqref{variation_closed} and \eqref{variation_4}, we compute
\eq{
  \label{variation_4_2}
  0&= \frac{1}{2}\:
  \frac{2\pi}{l_s^4}\int_{\mathbb{R}^{3,1}} \delta C_4\wedge\Biggl[
  \frac{1}{l_s^4}\int_{\mathcal{X}}\Bigl( d\widetilde F_5
    - H_3\wedge \widetilde F_3\Bigr)
    + \sum_{{\rm D}3_b,{\rm D}3_{b'}} N_{{\rm D}3_b} - 
    \sum_{{\rm O}7_i}\frac{N_{{\rm O}3_i}}{2} \\
  &\hspace{32pt}
  -\frac{1}{l_s^4}\sum_{{\rm D}7_a,{\rm D}7_{a'}}\int_{\Gamma_{{\rm D}7_a}}
    \left( \,\ch_2\bigl( \ov \F_a\bigr)  
    +l_s^4\,N_{{\rm D}7_a}\frac{c_2\bigl( \Gamma_{{\rm D}7_a}\bigr)}{24} \right)
    - \sum_{{\rm O}7_j}\int_{\Gamma_{{\rm O}7_j}} 
    \frac{c_2\bigl( \Gamma_{{\rm O}7_j}\bigr)}{6}
    \Biggr].
}

\bigskip
By the same arguments as for the D$5$-brane tadpole, the dependence of \eqref{variation_4_2} on $B_2^-$ should vanish. In order to see this, we employ the definition \eqref{def_ch} to obtain
\eq{
  \label{ch_2_exp}
  \ch_2\bigl( \ov \F \bigr) = 
  \ch_2\bigl( \ov \F^+ \bigr)
  + \ch_1\bigl( \ov \F^+ \bigr)\wedge\bigl(\varphi^* B_2^-\bigr)
  + \frac{N_{{\rm D}7}}{2} \: \bigl( \varphi^* B_2^- \bigr)^2
  \;,
}
which we use to calculate
\begin{align}
  \nonumber
  & \quad\,\sum_{{\rm D}7_a,{\rm D}7_{a'}}\int_{\Gamma_{{\rm D}7_a}}
  \ch_2\bigl( \ov \F_a\bigr) \\
  \nonumber
  &= \sum_{{\rm D}7_a,{\rm D}7_{a'}}\int_{\Gamma_{{\rm D}7_a}}\biggl[
  \ch_2\bigl( \ov \F^+_a\bigr)
  + \ch_1\bigl( \ov \F^+_a \bigr)\wedge\bigl(\varphi^* B_2^-\bigr)
  \biggr] \\
  &\nonumber\hspace{120.7pt}+ 
  \frac{1}{2}\sum_{{\rm D}7_a}\int_{\mathcal{X}}
  \bigl( B_2^- \bigr)^2 N_{{\rm D}7_a} \Bigl( [\Gamma_{{\rm D}7_a}]
  + [\Gamma_{{\rm D}7_a}'] \Bigr)
   \\[-3mm]
  &= \sum_{{\rm D}7_a,{\rm D}7_{a'}} \int_{\Gamma_{{\rm D}7_a}}
  \ch_2\bigl( \ov \F^+_a\bigr) \;.
\end{align}
In going from the second to the third line, we utilized the D$5$-brane tadpole cancellation condition to observe that the terms involving $\ch_1 (\ov \F^+)$ have to vanish, and for the cancellation of the expressions containing $(B_2^-)^2$, we used the same reasoning as in \eqref{tadpole_5_10}.

\bigskip
Next, following \cite{Collinucci:2008pf} (see also \cite{Braun:2008ua}), D$7$-branes on the orientifold space can have double-instersection points  and can therefore be singular. Thus, the definition of the corresponding Euler characteristic 
\eq{
  \label{usual_euler}
  \chi\bigl( \Gamma \bigr) = \int_{\Gamma} c_2\bigl( \Gamma \bigr)
}  
is ambiguous. 
However, as has been explained in \cite{Aluffi:2007sx,Collinucci:2008pf}, the Euler characteristic of an appropriate blow-up of the singularity minus the number of pinch-points leads to the correct result. We will denote the physical Euler characteristic of \cite{Aluffi:2007sx,Collinucci:2008pf} by $\chi_o(\Gamma)$, which reduces to the usual Euler characteristic \eqref{usual_euler}
for smooth D$7$-branes.

Employing equation \eqref{flux_number} for the background fluxes and denoting  the total number of D$3$-branes by $N_{{\rm D}3}$ as well as the total number of O$3$-planes by $N_{{\rm O}3}$, we deduce from \eqref{variation_4_2} the D$3$-brane tadpole cancellation condition to be of the form
\eq{
  \label{tadpole_4}
  \bom{
  \!\!
  N_{{\rm D}3} +2\, N_{\rm flux} = \frac{N_{{\rm O}3}}{4} 
  +\frac{1}{l_s^4}\sum_{{\rm D}7_a}\int_{\Gamma_{{\rm D}7_a}}
    \hspace{-13pt}
    \ch_2\bigl( \overline{\mathcal{F}}^+_a\bigr)
  +\sum_{{\rm D}7_a}N_{{\rm D}7_a}\:\frac{\chi_o( \Gamma_{{\rm D}7_a}\bigr)}{24} 
  +\sum_{{\rm O}7_j} \frac{\chi\bigl( \Gamma_{{\rm O}7_j}\bigr)}{12} .
  \!\!
  }
}


\section{Chiral Anomalies}
\label{sec_chiral_anom}

Before determining the chiral anomalies for a configuration of D$7$-branes, we are going to first comment on the possible gauge groups. To do so, let us denote the gauge group on a stack of $N_{{\rm D}7}$ D$7$-branes without gauge flux by $G$, which for type II constructions usually is $U(N_{{\rm D}7})$ or $Sp(2N_{{\rm D}7})$ respectively $SO(2N_{{\rm D}7})$. If we turn on gauge flux $\ov F$ with structure group $\ov H\subset G$, then the observable gauge group $H$ is the commutant of $\ov H$ in $G$
\eq{
  H = \Bigl\{ \,h\in G \::\: \bigl[ h, \ov h \bigr]=0 \quad\forall\;
  \ov h \in \ov H \,\Bigr\} \;. 
}
However,  in order to simplify our discussion, we will consider only $U(1)$ gauge fluxes on the D$7$-branes which are diagonally embedded into $U(N_{{\rm D}7})$ in the following way
\eq{
  \label{flux_choice}
  \ov F = \ov f\: \mathds{1}_{N_{{\rm D}7}\times N_{{\rm D}7}} \;.
}
Let us emphasize that the discussion in the following two sections of this paper relies on this choice of flux and its embedding. For a different structure group $\ov H$ or embedding into $G$, the calculations become slightly more involved.
\begin{table}[b!]
\centering
\renewcommand{\arraystretch}{1.5}
\arraycolsep8pt
\begin{equation*}
\begin{array}{c||c|c|c|c|c|c}
  & F & \ov F & S & \ov S & A & \ov A \\ \hline\hline
  \mbox{dim}(r) & N & N & \frac{N(N+1)}{2} & \frac{N(N+1)}{2}
    & \frac{N(N-1)}{2} & \frac{N(N-1)}{2} \\
  Q(r) & +1 & -1 & +2 & -2 & +2 & -2 \\
  C(r) & \frac{1}{2} & \frac{1}{2} & \frac{N+2}{2} & \frac{N+2}{2} 
    & \frac{N-2}{2} & \frac{N-2}{2} \\
  A(r) & +1 & -1 & N+4 & -(N+4) & N-4 & -(N-4) 
\end{array}
\end{equation*}
\caption{Group theoretical quantities for $SU(N)$ where $F$ stands for the fundamental, $S$ for the symmetric and $A$ for the anti-symmetric representation (see for instance \cite{MarchesanoBuznego:2003hp}). \label{table_group_quant}}
\end{table}  

\bigskip
We now turn to the chiral anomalies. The anomaly coefficients are expressed in terms of the cubic Casimir $A(r)$, the index $C(r)$ and the $U(1)$ charge $Q(r)$ where $r$ denotes a particular representation. For $SU(N)$, these quantities are summarized in table \ref{table_group_quant} and the discussion for $SO(2N)$ and $Sp(2N)$ gauge groups can be found in appendix \ref{app_sp_so}.
More concretely, the cubic non-abelian, the mixed abelian--non-abelian, the cubic abelian and the mixed abelian--gravitational anomalies are calculated via the following formulas (see for instance \cite{MarchesanoBuznego:2003hp})
\eq{
  \label{anom}
  \mathcal{A}_{SU(N_{{\rm D}7_a})^3} &= \sum_r A(r) \;, \\
  \mathcal{A}_{U(1)_a-SU(N_{{\rm D}7_b})^2} &= \sum_r Q_a(r) \, C_b(r) \;, \\
  \mathcal{A}_{U(1)_a-U(1)^2_b} &= \sum_r Q_a(r) \, Q^2_b(r) 
    \, \mbox{dim}(r)\;, \\
  \mathcal{A}_{U(1)_a-G^2} &= \sum_r Q_a(r)
    \, \mbox{dim}(r)\;.
}

\begin{table}[t!]
\begin{centering}
\renewcommand{\arraystretch}{1.5}
\begin{tabular}{c||c}
  Representation & Multiplicity \\ \hline\hline
  $(\ov{N}_a,N_b)$ & $I_{ab}$ \\
  $(N_a,N_b)$      & $I_{a'b}$ \\
  S$_a$   & $\frac{1}{2}( I_{a'a} - 2\, I_{O7a})$ \\
  A$_a$  & $\frac{1}{2}( I_{a'a} + 2\, I_{O7a})$ 
\end{tabular}
\caption{Formulas for determining the chiral spectrum. Here $(N_a,N_b)$ denotes a bi-fundamental representation of the gauge group $G_a\times G_b$ while $S_a$ and $A_a$ stand for the symmetric respectively anti-symmetric representation of the gauge group $G_a$.\label{table_spectrum}}
\end{centering}
\end{table}  

In order to determine these anomalies, we have to employ the rules for computing the chiral spectrum which are summarized in table \ref{table_spectrum}. 
The chiral index $I_{ab}$ used in this table is defined in the following way \cite{Minasian:1997mm,Katz:2002gh,Aspinwall:2004jr,Blumenhagen:2006ci,Marchesano:2007de}
\eq{
  \label{intersection_pre}
  I_{ab} = \int_{\mathcal{X}} \frac{1}{l_s^6}\left( 
  \frac{\ch_1\bigl( \varphi_*\ov \F_a\bigr)}{N_{{\rm D}7_a}}
  - \frac{\ch_1\bigl( \varphi_*\ov \F_b\bigr)}{N_{{\rm D}7_b}}
  \right) 
  \wedge \bigl[ \Gamma_{{\rm D}7_a}\bigr]
  \wedge \bigl[ \Gamma_{{\rm D}7_b}\bigr] \;.
}  
The somewhat unusual factors of $N_{{\rm D}7}^{-1}$ are due to the fact, that we are counting representations.\footnote{The chiral number of massless excitations between two D$7$-branes $a$ and $b$ is counted by the index $\widetilde I_{ab} = \int_{\mathcal{X}} \frac{1}{l_s^6}\Bigl( \ch_1\bigl( \varphi_*\ov \F_a\bigr)\ch_0\bigl( \varphi_*\ov \F_b\bigr)   - \ch_0\bigl( \varphi_*\ov \F_a\bigr) \ch_1\bigl( \varphi_*\ov \F_b\bigr)  \Bigr) 
\wedge \bigl[ \Gamma_{{\rm D}7_a}\bigr]
\wedge \bigl[ \Gamma_{{\rm D}7_b}\bigr] $, which in the present case reduces to \eqref{intersection_pre} when counting representations.
}
Next, employing our definitions \eqref{notation_10} and \eqref{notation_11}, we see that $B_2^-$ cancels out in \eqref{intersection_pre} and so, as expected, only the quantized flux $\ov\F^+$ contributes to the chiral index
\eq{
  \label{intersection}
  I_{ab} = \int_{\mathcal{X}} \frac{1}{l_s^6}\left( 
  \frac{\ch_1\bigl( \varphi_*\ov \F^+_a\bigr)}{N_{{\rm D}7_a}}
  - \frac{\ch_1\bigl( \varphi_*\ov \F^+_b\bigr)}{N_{{\rm D}7_b}}
  \right) 
  \wedge \bigl[ \Gamma_{{\rm D}7_a}\bigr]
  \wedge \bigl[ \Gamma_{{\rm D}7_b}\bigr] .
}


\subsubsection*{Cubic Non-Abelian Anomaly}

For the computation of the cubic non-abelian anomaly, we focus on the D$7$-brane labelled by $a$ and calculate using \eqref{intersection}
\eq{
   &\hspace{-32pt} \mathcal{A}_{SU(N_{{\rm D}7_a})^3}  \\[3mm]
   \hspace{25pt}
   &=     \sum_{{\rm D}7_b} N_{{\rm D}7_b}\, \bigl( I_{ba}+ I_{b'a} \bigr) 
   - 8 \,\sum_{{\rm O}7_i} I_{{\rm O}7_i\,a} \\
   &= -\int_{\mathcal{X}}  \frac{\ch_1\bigl( \varphi_*\ov \F^+_a\bigr)
     }{N_{{\rm D}7_a}}\wedge
     \left[ \Gamma_{{\rm D}7_a} \right]\wedge
     \biggl(  \sum_{{\rm D}7_b} N_{{\rm D}7_b}
     \Bigl(\, [\Gamma_{{\rm D}7_b}] +[\Gamma'_{{\rm D}7_b}]\,\Bigr) - 8 \sum_{{\rm O}7_i}\: 
     [\Gamma_{{\rm O}7_i}] \biggr) \\
   &\quad\, + \int_{\mathcal{X}}
   \sum_{{\rm D}7_b} \biggl(
    \ch_1\bigl(\varphi_*\ov\F^+_b\bigr) \wedge
    \left[ \Gamma_{{\rm D}7_b} \right]
    +  \ch_1\bigl(\varphi_*{\ov\F^+}'_b\bigr) \wedge 
    \left[ \Gamma'_{{\rm D}7_b} \right]  
    \biggr)\wedge  \left[ \Gamma_{{\rm D}7_a} \right] 
   .
   \label{calc_cub_nonab}
}
Here, the prime again denotes the $\Omega(-1)^{F_L}\sigma$ image and the sums run over all D$7$-branes $b$ respectively all O$7$-planes.
Employing  the D$7$-brane tadpole cancellation condition \eqref{tadpole_8}, we see that the first line in \eqref{calc_cub_nonab} vanishes. For the vanishing of the second line, we use the D$5$-brane tadpole cancellation condition \eqref{tadpole_6} to arrive at
\eq{
  \label{cub_non_ab}
  \mathcal{A}_{SU(N_{{\rm D}7_a})^3} =0 \;.
}


\vskip30pt
\subsubsection*{Mixed Abelian--Non-Abelian Anomaly}

Next, we consider the mixed abelian--non-abelian anomaly. Along the same lines as above, we compute
\begin{align}
   \nonumber
   \mathcal{A}_{U(1)_a-SU(N_{{\rm D}7_b})^2} &=  
   \frac{1}{2} \,\delta_{ab} \left(
   \sum_{{\rm D}7_c} N_{{\rm D}7_c} \bigl( I_{cb}+ I_{c'b} \bigr) - 8 \sum_{{\rm O}7_i}
   I_{{\rm O}7_i\,b} \right) 
   -\frac{N_{{\rm D}7_a}}{2}\, \Bigl( I_{ab} - I_{a'b} \Bigr) \\[1.6mm]
   \label{anom_mana}
   &= -\frac{N_{{\rm D}7_a}}{2}\, \Bigl( I_{ab} - I_{a'b} \Bigr) \;,
\end{align}
where we used, similarly as for the cubic non-abelian anomaly, the tadpole cancellation conditions \eqref{tadpole_8} and \eqref{tadpole_6} for the vanishing of the first term.


\subsubsection*{Cubic Abelian Anomaly}

For the cubic abelian anomaly, we find  
\begin{align}
   \nonumber
   &\quad \mathcal{A}_{U(1)_a-U(1)_b^2}  \\[1.25mm]
   \nonumber
   &=  
   \frac{N_{{\rm D}7_a}}{3} \:\delta_{ab} \left(
   \sum_{{\rm D}7_c} N_{{\rm D}7_c}\, \bigl( I_{cb}+ I_{c'b} \bigr) - 
   8 \sum_{{\rm O}7_i}
   I_{{\rm O}7_i\,b} \right) 
   -N_{{\rm D}7_a}\,N_{{\rm D}7_b}\: \Bigl( I_{ab} - I_{a'b} \Bigr) \\
   \label{anom_ca}
   &=  -N_{{\rm D}7_a}\,N_{{\rm D}7_b}\: \Bigl( I_{ab} - I_{a'b} \Bigr)
   \;,
\end{align}  
where the pre-factor $\frac{1}{3}$ is due to the additional symmetry in the case $a=b$, and we again used the tadpole cancellation conditions \eqref{tadpole_8} and \eqref{tadpole_6}.


\vskip10mm
\subsubsection*{Mixed Abelian--Gravitational Anomaly}

From \eqref{anom}, we finally determine the mixed abelian--gravitational anomaly of a D$7$-brane $a$. Employing the tadpole cancellation condition \eqref{tadpole_8}, we obtain
\begin{align}
   \nonumber
   \mathcal{A}_{U(1)_a-G^2} & =  
   N_{{\rm D}7_a} \left(
   \sum_{{\rm D}7_b} N_{{\rm D}7_b}\, \bigl( I_{ba}+ I_{b'a} \bigr) - 
   2\sum_{{\rm O}7_i}
   I_{{\rm O}7_i\,a} \right)  \\
   \nonumber
   & =  N_{{\rm D}7_a} \left(
   \sum_{{\rm D}7_b} N_{{\rm D}7_b}\, \bigl( I_{ba}+ I_{b'a} \bigr) - 8\sum_{{\rm O}7_i}
   I_{{\rm O}7_i\,a} \right) 
   + 6\,N_{{\rm D}7_a}  \sum_{{\rm O}7_i}  I_{{\rm O}7_i\,a}\\[1.6mm]
   \label{anom_grav}
   &= 6\,N_{{\rm D}7_a}  \sum_{{\rm O}7_i}  I_{{\rm O}7_i\,a} \;.
\end{align}



\section{The Generalized Green--Schwarz Mechanism}
\label{sec_gsm}

\begin{figure}[ht]
\vskip10pt
\begin{center}
  \hspace*{15pt}
  \subfigure[Anomaly Diagram]{
  \includegraphics[width=0.35\textwidth]
  {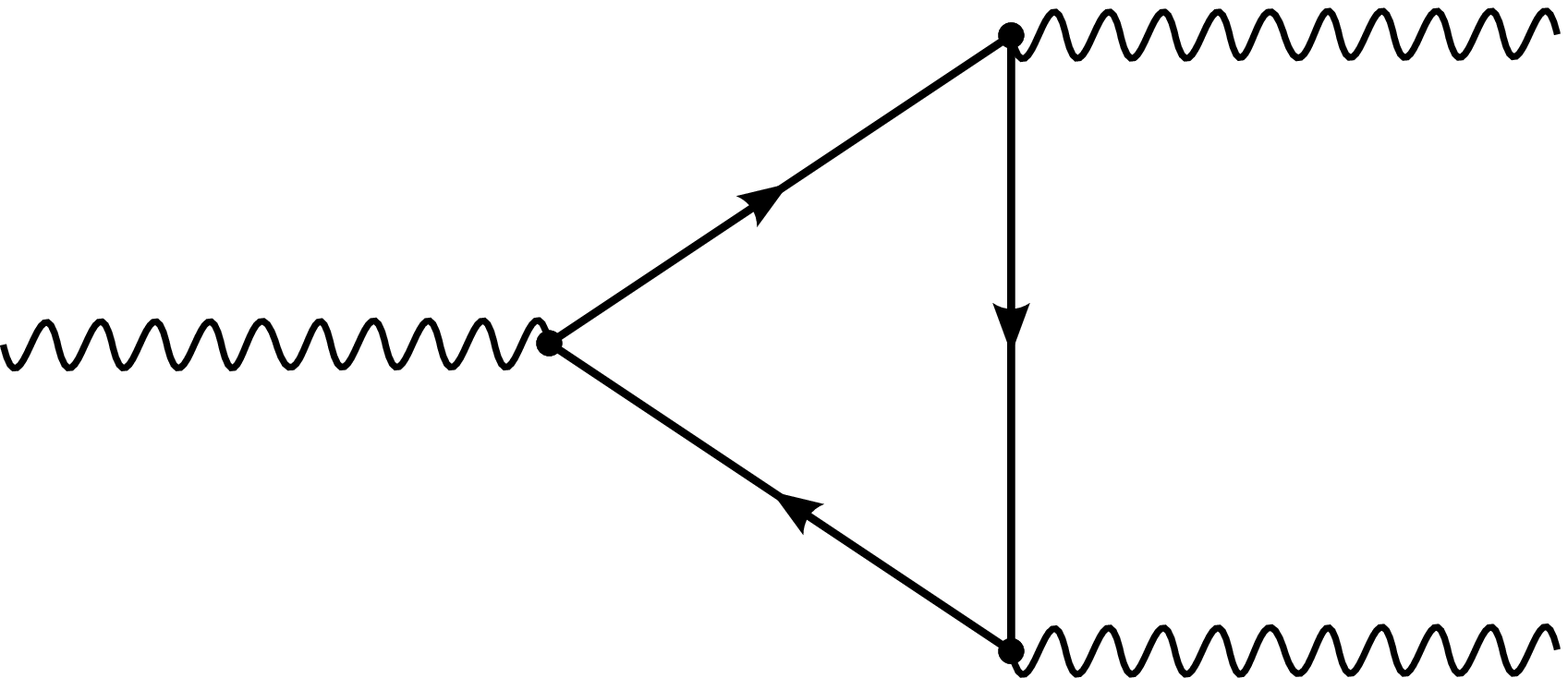}
  \begin{picture}(0,0)
   \put(1,1){$\scriptstyle F$}
   \put(1,58){$\scriptstyle F$}
   \put(-158,29){$\scriptstyle F$}
  \end{picture}
  \label{fig_anom}
  }
  \hfill
  \subfigure[Green--Schwarz Diagram]{
  \includegraphics[width=0.35\textwidth]{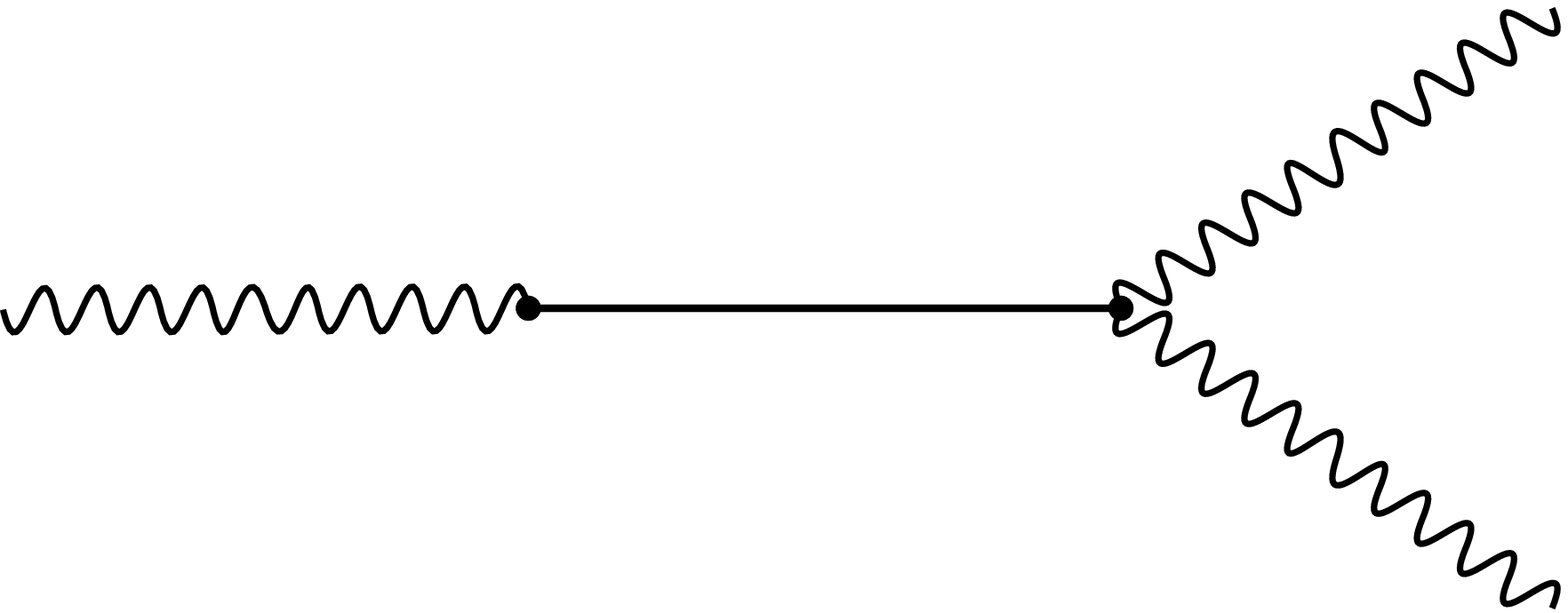}
  \begin{picture}(0,0)
   \put(0,-3){$\scriptstyle F$}
   \put(0,56){$\scriptstyle F$}
   \put(-159,27){$\scriptstyle F$}
   \put(-95,35){$\scriptstyle A_{(2)}$}
   \put(-63,20){$\scriptstyle B_{(0)}$}
  \end{picture}
  \label{fig_gsm}
  }
  \hspace*{15pt}
\end{center}
\vspace*{-15pt}
\caption{Anomaly and Green-Schwarz diagrams.}
\end{figure}

In type II string theory constructions with D-branes, 
chiral anomalies originating from a diagrams such as in figure \ref{fig_anom} are cancelled via the generalized Green--Schwarz mechanism 
\cite{Green:1984sg,Sagnotti:1992qw,Aldazabal:1998mr,Ibanez:1998qp,Aldazabal:1999nu,Ibanez:1999pw,Scrucca:1999zh,Aldazabal:2000dg}.
The key observation for this mechanism to work is that in four dimensions a two-form $A_{(2)}$ and a scalar $B_{(0)}$ are dual to each other via the Hodge-$\star$-operation
\eq{
  \label{gsm_class_1}
  dA_{(2)}\sim \star_4\, d B_{(0)} \;.
}
Then, if there are couplings in the four-dimensional action of the form
\eq{
  \label{gsm_class_2}
  \mbox{tr}\bigl( F \bigr)\, A_{(2)}
  \qquad\mbox{and}\qquad
  \mbox{tr}\bigl( F^2 \bigr)\, B_{(0)} \;,
}
one can construct diagrams which cancel the chiral anomalies. An example of such a Green-Schwarz diagram can be found in figure \ref{fig_gsm}. 


\subsection{Green--Schwarz Couplings}

In the present context, the two-forms $A_{(2)}$ and scalars $B_{(0)}$  are obtained from a dimensional reduction of the R-R $p$-form potentials $C_p$ and the duality \eqref{gsm_class_1} is provided by \eqref{duality}. 
To see this in more detail, we perform a dimensional reduction of $C_2$, $C_4$ and $C_6$ on the Calabi-Yau manifold $\mathcal{X}$
\eq{
  \label{dim_red_RR}
  \begin{array}{lcl@{}lll@{}llll@{}l}
    C_2 &=& \C^I& \omega_I &+& \D_0  \\
    C_4 &=& C_I & \sigma^I &+& D^I & \wedge\omega_I & +\ldots \\
    C_6 &=&     &          & & \D_I& \wedge\sigma^I & + \ldots 
  \end{array}
}  
where $C_I$ respectively $\C^I$ are four-dimensional scalars and $D^I$ as well as $\D_I$ are two-forms in $\mathbb{R}^{3,1}$. The ellipsis indicate that there are further terms coming for instance from \pagebreak[3] the reduction of $C_p$ on three-cycles and from the reduction on $\mathcal{X}$. However, these terms  will not be of relevance here. Let us also note that from \eqref{duality}, we obtain
\eq{
  \label{rel_sign}
  \arraycolsep1.75pt
  \begin{array}{lclclcr}
  d \C^I &=& - \star_4 d \D_I
    &\hspace{40pt}\Rightarrow\hspace{40pt}&
  \C^I &\leftrightarrow& -\D_I  \\[2mm]
  d C_I &=& + \star_4 d D^I
    &\hspace{40pt}\Rightarrow\hspace{40pt}&
  C_I &\leftrightarrow& +D^I 
  \end{array}
}
where $\star_4$ is the Hodge-$\star$-operator in four dimensions. The relative sign between these two dualities will be important in the following.

\bigskip
We now turn to the couplings \eqref{gsm_class_2} which are contained in the Chern-Simons actions of the D-branes and O-planes. To determine these, we expand the holomorphic divisor wrapped by a D$7$-brane $a$ as 
\eq{
  \Gamma_{{\rm D}7_a}=m_a^I \,\gamma_I \;,
  \hspace{50pt}
  m_a^I \in\mathbb{Z}\;,
}
where $\{\gamma_I\}$ is the basis of four-cycles introduced in \eqref{basis_4_cycles}. Next, since here we are considering gauge groups $U(N)$ for which the corresponding algebra satisfies $\mathfrak{u}(N)\simeq\mathfrak{u}(1)\times\mathfrak{su}(N)$, we write the four-dimensional open string field strength $F$ as
\eq{
  \label{decomp_field_strength}
  F = f \,\mathds{1} + \sum_A \mathsf{F}^A \,T^A\;,
}
where $f$ denotes the abelian and $\mathsf{F}^A$ stands for the non-abelian part. For the anti-symmetric representations matrices $T^A$ of the gauge group in the fundamental representation, we have
\eq{
  \label{traces}
  \mbox{tr}\, \bigl( T^A \bigr) = 0 \;,
  \hspace{40pt}
  \mbox{tr}\,\bigl( T^A T^B \bigr) = \frac{1}{2}\: \delta^{AB} \;,
}
where the latter relation reflects the usual choice of normalization. Using then \eqref{flux_choice} and \eqref{decomp_field_strength} together with \eqref{traces}, we can evaluate some quantities needed in the following
\eq{
  \label{chern_exp_1}
  \ch_1\bigl( \F \bigr) &= N \left( l_s^2\: \frac{f+ \ov f}{2\pi}
    +\varphi^* B_2 \right) \;, \\
  \ch_2\bigl( \F \bigr) &=\frac{1}{2}\left[
    \frac{l_s^4}{8\pi^2}\:\sum_A \mathsf{F}^A\mathsf{F}^A
    + N \left( l_s^2\: \frac{f+ \ov f}{2\pi}
    +\varphi^* B_2 \right)^2
    \right] \;, \\
  \ch_3\bigl( \F \bigr) &=\frac{1}{6}\left[
    \frac{3 l_s^4}{8\pi^2}\:\sum_A \mathsf{F}^A\mathsf{F}^A
    \left( l_s^2\: \frac{\ov f}{2\pi}
    +\varphi^* B_2 \right)
    + N \left( l_s^2\: \frac{f+ \ov f}{2\pi}
    +\varphi^* B_2 \right)^3
    \right] ,
}
where $\ov f$ was the $U(1)$ gauge flux on the D$7$-branes introduced in \eqref{flux_choice} and appropriate wedge products are understood.

\bigskip
Given these expressions, we can now identify the Green-Schwarz couplings. In particular, the $\mbox{tr}\bigl( F\bigr)$ terms are obtained from the D$7$-brane action and read
\eq{
  \label{coupl_1}
  \mathcal{S}^{\rm CS}_{{\rm D}7} & = 
  -\mu_7 \int_{{\rm D}7}
  \biggl[ \ch_1\bigl( \F \bigr)\wedge C_6
  + \ch_2\bigl( \F \bigr) \wedge C_4 \biggr]
   + \ldots \\
  &= -\frac{2\pi}{l_s^4} \int_{\mathbb{R}^{3,1}} 
  \frac{l_s^2}{2\pi}\: N_{{\rm D}7}\: f  \wedge\left[ 
  \D_I \,m^I
  + \frac{1}{l_s^4}\,D^I \wedge\int _{\Gamma_{{\rm D}7}} \!\!\left( 
  \frac{l_s^2}{2\pi}\:\ov f
  + \varphi^* B^+_2\right) \wedge
  \omega_I \right]  +\ldots
}
where the ellipsis denote further couplings not of importance here.
The relevant terms involving $\mbox{tr}\bigl( F^2\bigr)$ read
\begin{align}
  \nonumber
  \mathcal{S}^{\rm CS}_{{\rm D}7} & = 
  -\mu_7 \int_{{\rm D}7}
  \biggl[ \ch_2\bigl( \F \bigr)\wedge C_4
  + \ch_3\bigl( \F \bigr) \wedge C_2 \biggr]
   + \ldots \\
  \label{coupl_2}
  &= -\frac{2\pi}{l_s^4} \int_{\mathbb{R}^{3,1}} 
    \frac{1}{2} \left( \frac{l_s^2}{2\pi}\right)^2
    \left( \frac{1}{2}\:\sum_A \mathsf{F}^A\mathsf{F}^A
    + N_{{\rm D}7}\: f^2 
    \right) \wedge \\
  \nonumber
  &\hspace{120pt}
  \wedge\left[  C_I \,m^I + \frac{1}{l_s^4}\:
  \C^I \int_{\Gamma_{{\rm D}7}} 
  \left( \frac{l_s^2}{2\pi}\:\ov f
  + \varphi^* B^+_2\right) \wedge\omega_I \right] + \ldots \;.
\end{align}
The  $\mbox{tr}\bigl( R^2\bigr)$ couplings are contained in the D$7$-brane action and can be determined using \eqref{top_exp_1} to be of the following form
\begin{align}
  \nonumber
  \mathcal{S}^{\rm CS}_{{\rm D}7} & = 
  -\mu_7 \int_{{\rm D}7} \frac{1}{96} \left( \frac{l_s^2}{2\pi}\right)^2 
   \mbox{tr}\,\bigl( R^2 \bigr)\wedge
  \biggl( \ch_0\bigl( \F \bigr)\, C_4 
  + \ch_1\bigl( \F \bigr) \wedge C_2 \biggr)
   + \ldots \\
  \label{coupl_3}
  &= -\frac{2\pi}{l_s^4} \int_{\mathbb{R}^{3,1}} 
  \frac{1}{96} \left( \frac{l_s^2}{2\pi}\right)^2 
   \mbox{tr}\,\bigl( R^2 \bigr)
   \wedge \\
  \nonumber
  &\hspace{75pt}
  \wedge
  \left[ N_{{\rm D}7}\, C_I \,m^I  
  + \frac{1}{l_s^4}\:N_{{\rm D}7}\:
  \C^I \int_{\Gamma_{{\rm D}7}} 
  \left( \frac{l_s^2}{2\pi}\:\ov f
  + \varphi^* B^+_2\right) \wedge\omega_I \right]  + \ldots \,,
\end{align}
while from the O$7$-plane action, we infer the terms
\eq{
  \label{coupl_4}
  \mathcal{S}^{\rm CS}_{{\rm O}7} & = 
  -\mu_7 \,Q_7 \int_{{\rm O}7} 
  \left( - \frac{1}{192} \left( \frac{l_s^2}{2\pi}\right)^2 
   \mbox{tr}\,\bigl( R^2 \bigr) \right)
  \wedge C_4 
   + \ldots \\
  &= -\frac{2\pi}{l_s^4} \int_{\mathbb{R}^{3,1}} 
  \frac{1}{24} \left( \frac{l_s^2}{2\pi}\right)^2 
   \mbox{tr}\,\bigl( R^2 \bigr) 
   \: C_I \,m^I  
   + \ldots \;.
}
A summary of the  couplings relevant for the generalized Green--Schwarz mechanism in the present context can be found in table \ref{couplings_detail}, where we employed again the notion of Chern characters.

\begin{table}[ht]
\begin{center}
\renewcommand{\arraystretch}{2.25}
\arraycolsep3pt
\eq{
  \nonumber
  \begin{array}{||@{\hspace{10pt}}rcrl@{\hspace{10pt}}||}
  \hline\hline
  f_a - \D_I & \quad : \quad &
    \displaystyle \frac{l_s^2}{2\pi} \hspace{14pt} & 
    \displaystyle N_{{\rm D}7_a}
    \: m_a^I \;, \\
  f_a - D^I &:& 
    \displaystyle \frac{l_s^2}{2\pi} \hspace{14pt}& 
    \displaystyle 
    \frac{1}{l_s^4}\,\int _{\Gamma_{{\rm D}7_a}} \ch_1\bigl( \ov \F_a^+ \bigr)
    \wedge
     \omega_I  \;, \\
  f^2_a - \C^I &: & 
    \displaystyle \left(\frac{l_s^2}{2\pi}\right)^2 &
    \displaystyle \frac{1}{2} \:
    \frac{1}{l_s^4}\,\int _{\Gamma_{{\rm D}7_a}} \ch_1\bigl( \ov \F_a^+ \bigr)
    \wedge
     \omega_I  \;,   \\
  f^2_a - C_I  &: & 
    \displaystyle \left(\frac{l_s^2}{2\pi}\right)^2 &
    \displaystyle \frac{N_{{\rm D}7_a}}{2}
    \: m_a^I \;,     \\
  \mathsf{F}^2_a - \C^I & : &
    \displaystyle \left(\frac{l_s^2}{2\pi}\right)^2 &
    \displaystyle \frac{1}{4\: N_{{\rm D}7_a}} \:
    \frac{1}{l_s^4}\,\int _{\Gamma_{{\rm D}7_a}} \ch_1\bigl( \ov \F_a^+ \bigr)
    \wedge
     \omega_I  \;,   \\
  \mathsf{F}^2_a - C_I & : & 
    \displaystyle \left(\frac{l_s^2}{2\pi}\right)^2 &
    \displaystyle \frac{1}{4}
    \: m_a^I \;,     \\
  R^2 - \C^I &: & 
    \displaystyle \left(\frac{l_s^2}{2\pi}\right)^2 &
    \displaystyle \frac{1}{96}\:
    \frac{1}{l_s^4}
    \sum_{a,a'}\int _{\Gamma_{{\rm D}7_a}} \ch_1\bigl( \ov \F_a^+ \bigr)\wedge
     \omega_I   \;,   \\
  R^2 - C_I & : & 
    \displaystyle \left(\frac{l_s^2}{2\pi}\right)^2 &
    \displaystyle \frac{1}{96}\:
    \left( \sum_{a,a'} N_{{\rm D}7_a}\: m_a^I 
    + 4 \sum_{{\rm 
    O}7_i} m_{{\rm O}7_i}^I
    \right).    \\
    \hline \hline        
  \end{array}
}
\caption{Summary of couplings relevant for the generalized Green--Schwarz mechanism in the context of type IIB orientifolds with D$3$- and D$7$-branes. Note that in $\ov\F^+$ only the diagonally embedded $U(1)$ flux \eqref{flux_choice} is turned on.\label{couplings_detail}}
\end{center}
\end{table}



\subsection{Green--Schwarz Diagrams}

In this subsection, we compute the contribution of the Green--Schwarz diagrams to the chiral anomalies.


\subsubsection*{Cubic Non-Abelian Anomaly}

For the cubic non-abelian anomaly, we see that there are no couplings of the form $\mathsf{F}- \D_I$ or $\mathsf{F}- D^I$ contained in the Chern-Simons actions \eqref{action_cs}. We therefore cannot construct the corresponding Green-Schwarz diagrams and so
\eq{
  \mathcal{A}^{\rm GS}_{SU(N_{{\rm D}7})^3} = 0\;.
}
This is expected since the cubic non-abelian anomaly \eqref{cub_non_ab} vanishes due to the tadpole cancellation conditions and does not need to be cancelled.


\subsubsection*{Mixed Abelian--Non-Abelian Anomaly}

Next, we consider the mixed abelian--non-abelian anomaly. The schematic form of the diagrams to be  evaluated is
\eq{
  f_a - \D_I
  - \C^I
  - \mathsf{F}_b^2 &
  \;,\hspace{50pt}
  f_a - D^I
  - C_I
  - \mathsf{F}_b^2 \;, \\
  f_a - \D_I
  - \C^I
  - \mathsf{F}_{b'}^2\! &
  \;,\hspace{50pt}
  f_a - D^I
  - C_I
  - \mathsf{F}_{b'}^2\! \;,
}
and with the help of the couplings shown in table \ref{couplings_detail},
we compute 
\eq{
  \mathcal{A}^{\rm GS}_{U(1)_a-SU(N_{{\rm D}7_b})^2} 
  =& \quad\left( \frac{l_s^2}{2\pi} \right)^3 N_{{\rm D}7_a}\: m_a^I \:
    \bigl(-1\bigr)  \:
    \frac{1}{4\, N_{{\rm D}7_b}}\:\frac{1}{l_s^4}\int_{\Gamma_{{\rm D}7_b}}
    \!
    \ch_1\bigl( \ov \F_b^+\bigr)\wedge\omega_I \\
  & + \left( \frac{l_s^2}{2\pi} \right)^3 
    \frac{1}{l_s^4}\int_{\Gamma_{{\rm D}7_a}}
    \ch_1\bigl( \ov \F_a^+\bigr)\wedge\omega_I\:
    \bigl(+1\bigr)  \:
    \frac{1}{4}\: m_b^I
  \\[1.5mm]    
  &+ \bigl( \;b \; \rightarrow \; b'\; \bigr) \\[1.5mm]    
  \label{gs_term_mana}
  =& \:\frac{1}{2}\,\left( \frac{l_s^2}{2\pi} \right)^3 \,
    \frac{N_{{\rm D}7_a}}{2} 
    \: \Bigl( I_{ab}- I_{a'b} \Bigr) 
    \hspace{-10pt}
}
where we have used \eqref{intersection} as well as $m_a^I\,\omega_I=\bigl[\Gamma_{{\rm D}7_a}\bigr]$. We also utilized that
\eq{
  I_{ab'}=-I_{a'b}
}
which is verified by employing \eqref{int_oim} and noting that $\ov \F^+$ is odd under $\Omega (-1)^{F_L}$. Comparing finally the Green--Schwarz contribution \eqref{gs_term_mana} to the anomaly \eqref{anom_mana},  we see that up to a numerical prefactor, \eqref{gs_term_mana} cancels the mixed abelian--non-abelian anomaly.


\subsubsection*{Cubic Abelian Anomaly}

For the cubic abelian anomaly, we need to compute the following  Green--Schwarz diagrams
\eq{
  f_a - \D_I
  - \C^I
  - f_b^2 &
  \;,\hspace{50pt}
  f_a - D^I
  - C_I
  - f_b^2 \;, \\
  f_a - \D_I
  - \C^I
  - f_{b'}^2 &
  \;,\hspace{50pt}
  f_a - D^I
  - C_I
  - f_{b'}^2 \;.
}
Performing the same steps as for the mixed abelian--non-abelian anomaly, we arrive at
\eq{
  \mathcal{A}^{\rm GS}_{U(1)_a-U(1)_b^2} 
  = \frac{1}{2}\,\left( \frac{l_s^2}{2\pi} \right)^3 \,
    N_{{\rm D}7_a}\,N_{{\rm D}7_b}
    \: \Bigl( I_{ab}- I_{a'b} \Bigr) 
     \;,
}
and by comparing with \eqref{anom_ca}, we see that the Green-Schwarz contribution cancels the cubic abelian anomaly up to the same prefactor as for the mixed abelian--non-abelian anomaly.


\subsubsection*{Mixed Abelian--Gravitational Anomaly}

Finally, the mixed abelian--gravitational anomaly is computed schematically as
\eq{
  f_a - \D_I
  - \C^I
  - R^2 
  \;,\hspace{50pt}
  f_a - D^I
  - C_I
  - R^2 \;.
}
Utilizing the couplings shown in table \ref{couplings_detail} as well as the D$5$- and D$7$-brane tadpole cancellation conditions, we find
\begin{align}
  \nonumber
  \mathcal{A}^{\rm GS}_{U(1)_a-G^2} 
  &= -\left( \frac{l_s^2}{2\pi} \right)^3 \frac{1}{96}\:
  \Biggl[ N_{{\rm D}7_a}\sum_{b,b'} \,\frac{1}{l_s^4}\,\int_{\Gamma_{{\rm D}7_b}} 
  \ch_1\bigl( \ov \F_b^+\bigr)\wedge\bigl[ \Gamma_{{\rm D}7_a} \bigr] \\
  \nonumber
  &\hspace{13pt}
  -\frac{1}{l_s^4} \int_{\Gamma_{{\rm D}7_a}} 
  \!\!\ch_1\bigl( \ov \F_a^+ \bigr)\wedge
  \left( \sum_{{\rm D}7_b} N_{{\rm D}7_b} \Bigl( 
  \bigl[ \Gamma_{{\rm D}7_a} \bigr] + \bigl[ \Gamma'_{{\rm D}7_b} \bigr] \Bigr)   
  + 4 \sum_{{\rm O}7_i} \bigl[ \Gamma_{{\rm O}7_i} \bigr]  \right)
  \Biggr]
  \\
  \label{gs_term_agrav}
  &=-\left( \frac{l_s^2}{2\pi} \right)^3 \frac{N_{{\rm D}7_a}}{8}
  \sum_{{\rm O}7_i} I_{{\rm O}7_i\,a} \;.
\end{align}
By comparing with \eqref{anom_grav}, we see that up to a numerical prefactor, the contribution from the Green--Schwarz diagrams \eqref{gs_term_agrav} has the right form to cancel the mixed abelian--gravitational anomaly.


\subsection{Massive U(1)s and Fayet-Iliopoulos Terms}
\label{sec_mass_fi}

To conclude this section, let us  comment on massive $U(1)$ factors and Fayet-Iliopoulos terms.
Using the definition of Chern characters \eqref{def_ch}, from equation \eqref{coupl_1} we can determine the St\"uckelberg mass terms for the gauge bosons on the D$7$-branes to be of the following form
\begin{align}
  \nonumber
  \mathcal{S}_{\rm mass} &= 
  -\frac{1}{l_s^2} \int_{\mathbb{R}^{3,1}} 
  \sum_{a,a'}
  f_{{\rm D}7_a}\wedge \left(
  N_{{\rm D}7_a} \,m_a^I \,\D_I 
  + \frac{1}{l_s^4}\,D^I \wedge \int _{\Gamma_{{\rm D}7_a}} 
  \ch_1\bigl( \ov \F_{{\rm D}7_a}^+ \bigr) \wedge  \omega_I \right) \\
  \label{mass_terms}
  &= 
  -\frac{1}{l_s^2} \int_{\mathbb{R}^{3,1}} 
  \sum_{{\rm D}7_a}
  f_{{\rm D}7_a} \wedge\biggl(
  N_{{\rm D}7_a} \bigl( m_a^I - m_{a'}^I\bigr)\,\D_I \\
  \nonumber
  &\hspace{140pt}
  + \frac{1}{l_s^4}\,D^I \wedge\int _{\Gamma_{{\rm D}7_a}} 
  \ch_1\bigl( \ov \F_{{\rm D}7_a}^+ \bigr) \wedge
  \bigl( \omega_I + \sigma^*\omega_I\bigr)
  \biggr)
\end{align}
where in going from the first to the second line we employed that the gauge field is odd under $\Omega(-1)^{F_L}$ together with equation \eqref{int_oim}. Let us next define the following two mass matrices for the gauge fields on the D$7$-branes
\eq{
  \label{mass_matrices}
   M_{I_+a}  = 
  \frac{1}{l_s^4}\: \int _{\Gamma_{{\rm D}7_a}} \!\!
  \ch_1\bigl( \ov \F^+_a \bigr) 
  \wedge\bigl( \omega + \sigma^*\omega\bigr)_{I_+}
  \,,\hspace{20pt}
  M_{a}^{I_-}  = 
  N_{{\rm D}7_a}\: \bigl( m_a -m_{a'}\bigr)^{I_-} ,
}
with $I_+=1,\ldots,h^{1,1}_+$ and $I_-=1,\ldots,h^{1,1}_-$.
Then, the massless (linear combinations of) $U(1)$ gauge fields on the D$7$-branes are those which are in the kernel of the combined matrix
\eq{
  M_{Ia} = \left[
  \begin{array}{c}
  \displaystyle 
  M_{I_+a}  \\[2.5mm]
  M^{I_-}_{\hspace{9.5pt}a}
  \end{array}
  \right]  \;.
}
Along the same lines as for the D$7$-branes, for the gauge fields on the D$3$-branes we find  that due to the orientifold images, there are no mass terms
\eq{
  \mathcal{S}_{\rm mass} &= -\frac{1}{l_s^2} \int_{\mathbb{R}^{3,1}}
  \sum_{b,b'} f_{{\rm D}3_b} \wedge \D_0 \: N_{{\rm D}3_b} = 0 \;.
}

\bigskip
With the help of the mass matrices \eqref{mass_matrices}, we can also determine the Fayet-Iliopoulos terms for the D$7$-branes. 
To do so, we first recall the definition of the axion-dilaton $\tau$, the moduli $G^{I_-}$ and the K\"ahler moduli $T_{I_+}$ \cite{Grimm:2004uq,Grimm:2007hs}
\eq{
  &\tau = C_0 + i\, e^{-\phi} 
  \;,\hspace{70.5pt}
  G^{I_-} = \int_{\Sigma^{I_-}} \Bigl( C_2 + \tau\, B_2^- \Bigr)
  \;, \\[2mm]
  &T_{I_+} = \int_{\gamma_{I_+}} \left(\:
  \frac{1}{2}\:e^{-\phi} J^2 + i\, C_4 + i\, B_2^- \wedge C_2
  +\frac{i}{2}\: \tau\bigl( B_2^- \bigr)^2 \right) .
}  
Here, $\{\gamma_{I}\}\in H_{4}(\mathcal{X},\mathbb{Z})$ and $\{\Sigma^{I}\}\in H_{2}(\mathcal{X},\mathbb{Z})$ are the basis of four- respectively two-cycles introduced in equation \eqref{basis_4_cycles}.
The derivatives of the K\"ahler potential $\mathcal{K}$ with respect to 
$\tau$, 
$G^{I_-}$ and $T_{I_+}$ read (see for instance the appendix of \cite{Grimm:2007hs})
\eq{
  \label{kaehler_deriv}
  &\frac{\partial \mathcal{K}}{\partial \tau}
    = \frac{i}{2} \: \frac{e^{\phi} }{\mathcal{V}}
    \left( \mathcal{V} - \frac{1}{2}
    \int_{\mathcal{X}} \bigl( B_2^-\bigr)^2\wedge J \right) \;, \\[2.5mm]
  &\frac{\partial \mathcal{K}}{\partial G^{I_-}}
    = \frac{i}{2} \: \frac{e^{\phi} }{\mathcal{V}}
    \int_{\gamma_{I_-}} B_2^-\wedge J
  \;,\hspace{40pt}
  \frac{\partial \mathcal{K}}{\partial T_{I_+}}
    = \frac{i}{2} \: \frac{e^{\phi} }{\mathcal{V}}
    \int_{\Sigma^{I_+}} J \;,
}
where $\mathcal{V}$ denotes the overall volume of the compactification space $\mathcal{X}$. Observing finally that the mass matrices \eqref{mass_matrices} correspond to the holomorphic Killing vectors of the gauged isometry associated to $T_{I_+}$ and $G^{I_-}$, we can compute the Fayet-Iliopoulos terms as
\eq{
  \xi_a \:&\sim\: -i\,M_{I_+a} \: 
    \frac{\partial \mathcal{K}}{\partial T_{I_+}}
    -i\, M_{a}^{I_-} \: \frac{\partial \mathcal{K}}{\partial G^{I_-}} \\
  \:&\sim\: \frac{1}{l_s^4}\: \frac{e^{\phi} }{\mathcal{V}}
    \int _{\Gamma_{{\rm D}7_a}} \ch_1\bigl( \ov \F^+_a \bigr) \wedge J
    + \frac{1}{2\,l_s^4}\: \frac{e^{\phi} }{\mathcal{V}}
    \int _{\Gamma_{{\rm D}7_a}-\Gamma'_{{\rm D}7_a}} N_{{\rm D}7_a}\: B_2^-\wedge J \\
  \:&\sim\: \frac{1}{l_s^4}\: \frac{e^{\phi} }{\mathcal{V}}
    \int _{\Gamma_{{\rm D}7_a}} \ch_1\bigl( \ov \F_a \bigr)\wedge J \;,
}
where we employed the definition \eqref{def_ch} as well as \eqref{int_oim} together with \eqref{orient_choice}. 
The vanishing of the D-term corresponding to a D$7$-brane without matter fields translates into $\xi_a=0$, which leads the well-known condition $\ov f\wedge J|_{\Gamma_{{\rm D}7_a}}=0$ for a  D$7$-brane with $U(1)$ flux $\ov f$ to be supersymmetric \cite{Dasgupta:1999ss,Marino:1999af,Jockers:2005zy}.


\section[Generalizations: Including D9- and D5-Branes]{Generalizations: D9- and D5-Branes}
\label{sec_generalization}

So far, we have studied the tadpole cancellation conditions and the generalized Green-Schwarz mechanism for type IIB orientifolds with D$3$- and D$7$-branes. However, it is possible to introduce also D$9$- and D$5$-branes which modify the tadpole cancellation conditions and therefore also the discussion for the chiral anomalies.

The reason for usually not considering D$9$- and D$5$-branes is that the orientifold projection maps them to anti-D$9$- and anti-D$5$-branes which are supersymmetric only at a particular point in moduli space.
Nevertheless, we can study the tadpole cancellation conditions and the chiral anomalies for such D-brane setups which we will do in some detail in this section.


\subsection{Tadpole Cancellation Conditions}

In order to determine the tadpole cancellation conditions,  let us recall equation \eqref{oaction_manif} and be more concrete about how the orientifold projection acts on the manifold a D$9$- or D$5$-brane is wrapping. In particular, we find
\eq{
  \label{orient_im_2}
  \Gamma_{{\rm D}9}' = - \Gamma_{{\rm D}9} \;, 
  \hspace{40pt}
  \Gamma_{{\rm D}5}' =
    - \sigma\, \Gamma_{{\rm D}5} \;,
}
where $\Gamma_{{\rm D}9}=\mathcal{X}$ is invariant under the holomorphic involution $\sigma$ and $\Gamma_{{\rm D}5}$ is a two-cycle in  $\mathcal{X}$ wrapped by a the D$5$-brane. Furthermore, note that we are also allowed  to turn on gauge flux $\ov F_{{\rm D}9}$ and $\ov F_{{\rm D}5}$ on the D$9$- respectively D$5$-branes which is odd under $\Omega(-1)^{F_L}$.


\subsubsection*{D9-Brane Tadpole Cancellation Condition}

In a very similar way as in section \ref{sec_tadpole}, we can now compute the D$9$-brane tadpole cancellation condition. The variation of the Chern-Simons action \eqref{action_cs} with respect to $C_{10}$ reads
\eq{
  \delta_{C_{10}}\mathcal{S}^{\rm CS}_{{\rm D}9} = 
  -\frac{2\pi}{l_s^{10}}\:\kappa_9 \int_{\mathbb{R}^{3,1}\times\mathcal{X}}
  \delta C_{10}\wedge \bigl[\Gamma_{{\rm D}9}\bigr] \wedge
  \ch_0\bigl( \ov \F_{{\rm D}9} \bigr) \;,
}  
where $[\Gamma_{{\rm D}9}]$ is the Poincar\'e dual of $\Gamma_{{\rm D}9}$ in $\mathcal{X}$, which is a zero-form, and the sign $\kappa_9$ had been introduced in equation \eqref{tension}.
Denoting the total number of D$9$-branes with gauge flux $F_a$ by $N_{{\rm D}9_a}=\ch_0( \ov\F_{{\rm D}9_a})$, we find for the equation of motion originating from $C_{10}$ that
\eq{
  \label{tadpole_10}
  \bom{
  0 = \kappa_9\sum_{{\rm D}9_a} N_{{\rm D}9_a}\:\Bigl( 
  \bigl[\Gamma_{{\rm D}9_a}\bigr] + 
  \bigl[\Gamma'_{{\rm D}9_a}\bigr] \Bigr) \;.
  }
}


\subsubsection*{D7-Brane Tadpole Cancellation Condition}

For the D$7$-brane tadpole cancellation condition, we compute the variation of the D$9$-brane Chern-Simons action with respect to $C_8$ as
\eq{
  \delta_{C_{8}}\mathcal{S}^{\rm CS}_{{\rm D}9} = 
  &-\frac{2\pi}{l_s^{10}}\: \kappa_9 
  \int_{\mathbb{R}^{3,1}\times\mathcal{X}} \delta C_8\wedge
  \bigl[\Gamma_{{\rm D}9}\bigr]\wedge
  \ch_1\bigl( \ov \F_{{\rm D}9}\bigr) \;.
  \label{tadpole_8_mod_pre}
}
Taking into account the orientifold images and combining  \eqref{tadpole_8_mod_pre} with the variations of the D$7$-brane,  O$7$-plane and bulk action \eqref{variation_8} respectively \eqref{variation_closed}, we find the following tadpole cancellation condition
\eq{
  \label{tadpole_8_mod}
  \bom{
  \kappa_7\sum_{a,a'} N_{{\rm D}7_a}\:[\Gamma_{{\rm D}7_a}]
   +\kappa_9\sum_{b,b'} \bigl[\Gamma_{{\rm D}9_b}\bigr] 
   \wedge  \ch_1\bigl( \ov \F_{{\rm D}9_b}\bigr)
  =  8\, \kappa_7\sum_{{\rm O}7_i}\: 
  [\Gamma_{{\rm O}7_i}]
  }
}
where the prime  denotes the image under the orientifold projection $\Omega(-1)^{F_L} \sigma$.
However, in its present form \eqref{tadpole_8_mod} still depends on the continuous fields $B_2^-$ which is not desirable. But, writing out the first Chern character as 
\eq{
  \label{ch_1_B2_2}
  \ch_1\bigl(\ov\F_{{\rm D}9}\bigr) 
   = \ch_1\Bigl(  \ov\F^+_{{\rm D}9} \Bigr) 
   + N_{{\rm D}9}\, B_2^- \;,
}
and noting that $B_2^-$ is even under $\Omega(-1)^{F_L}\sigma$ while $[\Gamma_{{\rm D}9}]$ is odd, we see that the dependence of \eqref{tadpole_8_mod} on $B_2^-$ vanishes. We can thus simply replace $\ov\F\to\ov\F^+$ in the D$7$-brane tadpole cancellation condition above.


\subsubsection*{D5-Brane Tadpole Cancellation Condition}

Let us continue with the equation of motion for $C_6$.
The variation of the D$9$-brane Chern-Simons action is computed as
\eq{
  \delta_{C_{6}}\mathcal{S}^{\rm CS}_{{\rm D}9} = 
  &-\frac{2\pi}{l_s^{10}}\: \kappa_9 
  \int_{\mathbb{R}^{3,1}\times\mathcal{X}} 
  \delta C_6 \wedge \bigl[\Gamma_{{\rm D}9}\bigr] \wedge\left(
  \ch_2\bigl( \ov \F_{{\rm D}9}\bigr) + l_s^4 \:N_{{\rm D}9}\:
  \frac{c_2\bigl( \mathcal{X} \bigr)}{24} \right) \;,
  \label{tadpole_6_d9}
}
where we observed that the tangential bundle of a D$9$-brane is equal to the tangential bundle of $\mathcal{X}$. The contribution of a D$5$-brane to the equation of motion of $C_6$ is found to be
\eq{
  \label{tadpole_6_d5}   
   \delta_{C_{6}}\mathcal{S}^{\rm CS}_{{\rm D}5} = 
   -\frac{2\pi}{l_s^{10}} \:\kappa_5
   \int_{\mathbb{R}^{3,1}\times \mathcal{X}}
   \delta C_6 \wedge \bigl[ \Gamma_{{\rm D}5} \bigr]\wedge
   \:\ch_0 \bigl( \ov\F_{{\rm D}5} \bigr) \;,
}
where $[\Gamma_{{\rm D}5}]$ denotes the Poincar\'e dual of $\Gamma_{{\rm D}5}$ in $\mathcal{X}$.
Taking into account the orientifold images and combining \eqref{tadpole_6_d9} as well as \eqref{tadpole_6_d5} with the variations computed in \eqref{variation_6} and \eqref{variation_closed}, we arrive at
\eq{  
  \label{tadpole_6_mod}
  \bom{
  \begin{split}
  0= \int_{\mathcal{X}} \omega_I\wedge\Biggl[ \hspace{20pt} \kappa_7&
  \sum_{a,a'} \:
    \bigl[ \Gamma_{{\rm D}7_a} \bigr] \wedge 
    \ch_1\bigl(\varphi_*\ov\F_{{\rm D}7_a}\bigr)  \\
    +\kappa_9&\sum_{b,b'} \:
      \bigl[ \Gamma_{{\rm D}9_b} \bigr]\wedge\biggl(
      \ch_2\bigl( \ov \F_{{\rm D}9_b}\bigr) 
      + l_s^4 \:N_{{\rm D}9_b}\:\frac{c_2\bigl( \mathcal{X} \bigr)}{24}
     \biggr) \\
    +\kappa_5&\sum_{c,c'} 
    \: \bigl[\Gamma_{{\rm D}5_c}\bigr]
    \: N_{{\rm D}5_c} \quad  \Biggr] 
  \end{split}
  }
}
where $\{\omega_I\}$ is again a basis of $(1,1)$-forms on $\mathcal{X}$.
Since \eqref{tadpole_6_mod} still depends on $B_2^-$, let us employ \eqref{ch_1_B2} to separate out the $B_2^-$ part from the first Chern character and use the definition \eqref{def_ch} to write the second Chern character as
\eq{  
  \label{ch_2_B2_2}
  \ch_2\bigl( \ov \F_{{\rm D}9} \bigr) = 
  \ch_2\bigl( \ov \F^+_{{\rm D}9} \bigr)
  + \ch_1\bigl( \ov \F^+_{{\rm D}9} \bigr)\wedge B_2^-
  + \frac{N_{{\rm D}9}}{2} \: \bigl( B_2^- \bigr)^2 \;.
}
Utilizing then the  D$7$-brane tadpole  condition \eqref{tadpole_8_mod}, we see that the dependence of \eqref{tadpole_6_mod} on $B_2^-$ vanishes, and so we can simply replace $\ov\F\to\ov\F^+$ in \eqref{tadpole_6_mod}.


\vskip15mm
\subsubsection*{D3-Brane Tadpole Cancellation Condition}

To finish our discussion of the tadpole cancellation conditions, let us turn to the equation of motion for $C_4$. The variation of the D$9$-brane Chern-Simons action is calculated as
\eq{
  \delta_{C_4} \mathcal{S}_{{\rm D}9}^{\rm CS}
  = 
  -\frac{2\pi}{l_s^{10}}\:\kappa_9 \int_{\mathbb{R}^{3,1}\times
  \mathcal{X}} \delta C_4 
  \wedge \bigl[ \Gamma_{{\rm D}9}\bigr]\wedge \biggl(
  \ch_3\bigl( \ov \F_{{\rm D}9}\bigr) 
  + l_s^4 \:
  \frac{c_2\bigl( \mathcal{X} \bigr)}{24} \wedge
  \ch_1 \bigl( \ov \F_{{\rm D}9}\bigr) \biggr) ,
}
while for the D$5$-brane action we find
\eq{
  \delta_{C_4} \mathcal{S}_{{\rm D}5}^{\rm CS}
  =
  -\frac{2\pi}{l_s^{10}}\:\kappa_5 \int_{\mathbb{R}^{3,1}\times
  \mathcal{X}} \delta C_4 \wedge
  \bigl[ \Gamma_{{\rm D}5} \bigr] \wedge
  \ch_1\bigl( \varphi_*\ov \F_{{\rm D}5}\bigr) \;.
}
Taking into account the orientifold images and combining the two expressions above with \eqref{variation_4} as well as \eqref{variation_closed}, we arrive at

\eq{
  \label{tadpole_mod_4}
  \bom{
  \begin{split}
  4\,N_{\rm flux} = &\quad\:
    \frac{\kappa_7}{l_s^6}\sum_{a,a'}\int_{\mathcal{X}}
    \bigl[ \Gamma_{{\rm D}7_a} \bigr]\wedge 
    \biggl(
    \ch_2\bigl( \varphi_*\overline{\mathcal{F}}_{{\rm D}7_a}\bigr) 
    + l_s^4\:N_{{\rm D}7_a}\, \frac{c_2\bigl( \Gamma_{{\rm D}7_a}
    \bigr)}{24}\,
     \biggr) \\
  & +\frac{\kappa_7}{l_s^6}\sum_{{\rm O}7_i}\int_{\mathcal{X}}
    \bigl[ \Gamma_{{\rm O}7_i} \bigr]\wedge 
    \biggl( l_s^4\: \frac{c_2\bigl( \Gamma_{{\rm O}7_i} \bigr)}{6} \,
    \biggr)
  \\
  &+\frac{\kappa_9}{l_s^6}\sum_{b,b'} \int_{\mathcal{X}}
  \bigl[\Gamma_{{\rm D}9_b} \bigr]\wedge
  \biggl(
  \ch_3\bigl( \ov \F_{{\rm D}9_b}\bigr) 
   + l_s^4 \:\ch_1 \bigl( \ov \F_{{\rm D}9_b}\bigr)\wedge
  \frac{c_2\bigl( \mathcal{X} \bigr)}{24} \,
  \biggr) \\
  &
  + \frac{\kappa_5}{l_s^6} \sum_{c,c'} \int_{\mathcal{X}}
  \bigl[ \Gamma_{{\rm D}5_c} \bigr]  \wedge 
  \ch_1\bigl( \varphi_*\ov \F_{{\rm D}5_c}\bigr) \\
  &
  + \frac{\kappa_3}{l_s^6} \sum_{d,d'} \int_{\mathcal{X}}
  \bigl[ \Gamma_{{\rm D}3_d} \bigr]  \: N_{{\rm D}3_d} \\
  &
  + \frac{\kappa_3}{l_s^6} \sum_{{\rm O}3_j} \int_{\mathcal{X}}
  \bigl[ \Gamma_{{\rm O}3_j} \bigr]  \left( -\frac{1}{2} \right)
  \end{split}
  }
}
where $[\Gamma_{{\rm D}3}]=\mathcal{X}$ denotes the Poincar\'e dual of $\Gamma_{{\rm D}3}$ in $\mathcal{X}$. Note that we have organized the appearing terms for later convenience. 
Similarly as in the previous cases, the dependence of this tadpole cancellation condition on the continuous fields $B_2^-$ should vanish. And indeed, using the definition \eqref{def_ch}, we can write the third Chern character as
\eq{  
  \label{ch_3_B2}
  \ch_3\bigl( \ov \F_{{\rm D}9} \bigr) = 
  \ch_3\bigl( \ov \F^+_{{\rm D}9} \bigr)
  + \ch_2\bigl( \ov \F^+_{{\rm D}9} \bigr)\wedge B_2^-
  + \frac{1}{2}\:\ch_1\bigl( \ov \F^+_{{\rm D}9} \bigr)\wedge 
    \bigl(B_2^-\bigr)^2
  + \frac{N_{{\rm D}9}}{3!} \: \bigl( B_2^- \bigr)^3 \;,
}
while for the first and second Chern character we use \eqref{ch_1_B2_2} respectively \eqref{ch_2_B2_2}. The terms in \eqref{tadpole_mod_4} involving $B_2^-$ can then be summarized as
\eq{
  &\frac{1}{l_s^6}\,\sum_{a,a'}\int_{\mathcal{X}}
    B_2^-\wedge\bigl[ \Gamma_{{\rm D}7_a} \bigr]
    \wedge \ch_1\bigl( \varphi_*\overline{\mathcal{F}}^+_{{\rm D}7_a}\bigr) 
  \\
  +&\frac{\kappa_9}{l_s^6}\sum_{b,b'} \int_{\mathcal{X}}
  B_2^-\wedge\bigl[ \Gamma_{{\rm D}9_b} \bigr]\wedge\biggl(
  \ch_2\bigl( \ov \F^+_{{\rm D}9_b}\bigr) 
  + l_s^4 \:N_{{\rm D}9_b}\:
  \frac{c_2\bigl( \mathcal{X} \bigr)}{24} 
  \biggr) \\
  +&
  \frac{\kappa_5}{l_s^6} \sum_{c,c'} \int_{\mathcal{X}}
  B_2^-\wedge \bigl[ \Gamma_{{\rm D}5_c} \bigr]\;
  N_{{\rm D}5_c}    \;,
}
which cancel due to the  D$5$-brane tadpole cancellation condition \eqref{tadpole_6_mod}. In a very similar way as in section \ref{sec_tadpole}, we see that the terms in \eqref{tadpole_mod_4} proportional to  $(B_2^-)^2$ have to vanish due to the  D$7$-brane tadpole cancellation condition \eqref{tadpole_8_mod}. Finally, using \eqref{ch_3_B2}, we observe that the terms proportional to $(B_2^-)^3$ vanish, due to the tadpole cancellation condition \eqref{tadpole_10}. In \eqref{tadpole_mod_4}, we can therefore replace $\ov\F\to\ov\F^+$.


\subsection{Chiral Spectrum}

After having explicitly determined the tadpole cancellation conditions for a combined system of D$9$-, D$7$-, D$5$- and D$3$-branes, we will now formulate them in a more compact way. This will allow us to infer  the rules for determining the chiral spectrum from the vanishing of the cubic non-abelian anomaly more easily.


\subsubsection*{Summary of Tadpole Cancellation Conditions}

In order to express the tadpole cancellation conditions of the last subsection in a unified way, following for instance \cite{Aspinwall:2004jr} 
(see also \cite{Green:1996dd,Cheung:1997az,Minasian:1997mm}), we define the charges \footnote{Note that we actually have to formulate these expressions in terms of sheaves. A naive way to compensate for this inaccuracy is to evaluate the Chern characters on the submanifold they are defined on whenever possible.}
\eq{
  \label{top_charges}
  \mathcal{Q}\bigl(\,\Gamma_{{\rm D}p}\,, \ov\F^+_{{\rm D}p} \,\bigr) &= 
  \kappa_p\:\bigl[ \Gamma_{{\rm D}p} \bigr]\wedge
  \ch\bigl( \varphi_*\ov \F^+_{{\rm D}p} \bigr) \wedge
  \sqrt{\frac{\A\bigl( \R_{T_{{\rm D}p}} \bigr)}
  {\A\bigl( \R_{N_{{\rm D}p}} \bigr)}} \;, \\
  \mathcal{Q}\bigl(\,\Gamma_{{\rm O}p} \,\bigr) &= 
  \kappa_p\:Q_p\:\bigl[ \Gamma_{{\rm D}p} \bigr]\wedge
  \sqrt{\frac{\mathcal{L}\bigl( \R_{T_{{\rm D}p}}/4 \bigr)}
  {\mathcal{L}\bigl( \R_{N_{{\rm D}p}}/4 \bigr)}} \;.  
}
The quantities involved here had been introduced around equation \eqref{action_cs}, but let us note once more that $[\Gamma]$ denotes the Poincar\'e dual of $\Gamma$ in $\mathcal{X}$, the R-R charge of the O-planes was $Q_p=-2^{p-4}$ and  that the signs $\kappa_p=\pm 1$ had been introduced in \eqref{tension}.

By comparing the charges \eqref{top_charges} with the explicit tadpole cancellation conditions \eqref{tadpole_10}, \eqref{tadpole_8_mod} and \eqref{tadpole_6_mod}, we observe that the D$p$-brane tadpoles for $p=9,7,5$ can be expressed in the following way
\eq{
  \label{tadpole_compact_1}
       0   = \sum_{{\rm D}q,{\rm D}q'} 
  \mathcal{Q}\bigl(\,\Gamma_{{\rm D}q}\,, \ov\F^+_{{\rm D}q} \,\bigr)
  + \sum_{{\rm O}q}  
  \mathcal{Q}\bigl(\,\Gamma_{{\rm O}q}\, \bigr)
  \biggr|_{(9-p)-{\rm form}} \;.
}
In \eqref{tadpole_compact_1}, the restrictions selects to the zero-, two- and four-form part, and the sums in this and the following formulas run over all D$p$-branes as well as over all O$p$-planes. Concretely, this means
\eq{
  \label{expl_sums}
  \sum_{{\rm D}q,{\rm D}q'} =
  \sum_{{\rm D}9_a}+  \sum_{{\rm D}9_{a'}} +
  \sum_{{\rm D}7_b}+ \ldots +  \sum_{{\rm D}3_{d'}} 
  \;,\hspace{50pt}
  \sum_{{\rm O}q} =
  \sum_{{\rm O}7_i}+  \sum_{{\rm O}3_{j}} \;.
}
By comparing the charges \eqref{top_charges} with the explicit form of the D$3$-brane tadpole  \eqref{tadpole_mod_4}, we see that, using \eqref{flux_number}, this condition can be expressed as
\eq{
  \label{tadpole_compact_2}
       - H_3\wedge F_3   = \sum_{{\rm D}q,{\rm D}q'} 
  \mathcal{Q}\bigl(\,\Gamma_{{\rm D}q}\,, \ov\F^+_{{\rm D}q} \,\bigr)
  + \sum_{{\rm O}q} 
  \mathcal{Q}\bigl(\,\Gamma_{{\rm O}q}\, \bigr)
  \biggr|_{6-{\rm form}} \;.
}


\subsubsection*{Rules for Determining the Chiral Spectrum}

Let us now state the rules for computing the chiral spectrum in the present context. These have been inferred from the requirement that the cubic non-abelian anomaly should vanish using the tadpole cancellation condition.
For that purpose, following for instance \cite{Aspinwall:2004jr}, we define
\eq{
  I_{{\rm D}p\,{\rm D}q}
  &= \frac{1}{N_{{\rm D}p}\, N_{{\rm D}q} }
  \int_{\mathcal{X}}
  \mathcal{Q}\bigl(\,\Gamma_{{\rm D}p}\,, \ov\F^+_{{\rm D}p} \,\bigr)
  \wedge
  \mathcal{Q}\bigl(\,\Gamma_{{\rm D}q}\,, -\ov\F^+_{{\rm D}q} \,\bigr) \;, 
  \\[2mm]
  I_{{\rm O}p\,{\rm D}q}
  &= \frac{1}{N_{{\rm D}q} }
  \int_{\mathcal{X}}
  \mathcal{Q}\bigl(\,\Gamma_{{\rm O}p}\,\bigr)
  \wedge
  \mathcal{Q}\bigl(\,\Gamma_{{\rm D}q}\,, -\ov\F^+_{{\rm D}q} \,\bigr)
    \;.
}
Note that here the prefactor is again due to the fact that we are counting representations instead of the chiral number of massless excitations. The multiplicities of the bi-fundamental and the symmetric as well as anti-symmetric representations in terms of these indices are given in table \ref{rule_spec_gen}.
\begin{table}[t]
\begin{equation*}  
  \renewcommand{\arraystretch}{1.5}
  \arraycolsep8pt
  \begin{array}{c||c}
  \mbox{Representation} & \mbox{Multiplicity} \\ \hline\hline
  \displaystyle \bigl( \ov N_{{\rm D}p},  N_{{\rm D}q} \bigr) &
    \displaystyle I_{{\rm D}p\,{\rm D}q} \\
  \displaystyle \bigl( N_{{\rm D}p},  N_{{\rm D}q} \bigr) &
    \displaystyle I_{{\rm D}p'\,{\rm D}q} \\
  \displaystyle S_{{\rm D}p} & 
    \displaystyle {\textstyle \frac{1}{2}}\Bigl( \,
    I_{{\rm D}p'{\rm D}p} + {\textstyle \frac{1}{4}}\sum_{{\rm O}q} 
    I_{{\rm O}q{\rm D}p} \,\Bigr) \\
  \displaystyle A_{{\rm D}p} & 
    \displaystyle {\textstyle \frac{1}{2}}\Bigl( \,
    I_{{\rm D}p'{\rm D}p} - {\textstyle \frac{1}{4}}\sum_{{\rm O}q} 
    I_{{\rm O}q{\rm D}p} \,\Bigr) 
  \end{array}
\end{equation*}
\caption{Rules for determining the chiral spectrum for a combined system of D$9$-, D$7$-, D$5$- and D$3$-branes in the context of type IIB orientifolds with O$7$- and O$3$-planes. The sums run over all O-planes as in equation \eqref{expl_sums}.\label{rule_spec_gen}}
\end{table}


\subsubsection*{Chiral Anomalies}

The expressions in table  \ref{rule_spec_gen} had been adjusted to the fact that the generalized Green-Schwarz mechanism does not provide any terms to cancel the cubic non-abelian anomaly. This anomaly therefore has to vanish due to the tadpole cancellation conditions which we verify now. In particular, using \eqref{anom} and table \ref{table_group_quant}, we compute
\begin{align}
  \nonumber
  \mathcal{A}_{SU(N_{{\rm D}p})^3} =&
  \sum_{{\rm D}q \neq {\rm D}p} N_{{\rm D}q}
  \Bigl( I_{{\rm D}q\,{\rm D}p} + 
  I_{{\rm D}q'\,{\rm D}p} \Bigr) 
   + \bigl( N_{{\rm D}p}+4\bigr) 
  \:\frac{1}{2}\:\Bigl( \,
    I_{{\rm D}p'{\rm D}p} + {\textstyle \frac{1}{4}}\sum_{{\rm O}q} 
    I_{{\rm O}q\,{\rm D}p} \,\Bigr)\\
  \nonumber
  & \hspace{140.25pt}+ \bigl( N_{{\rm D}p}-4\bigr) 
  \:\frac{1}{2}\:\Bigl( \,
    I_{{\rm D}p'{\rm D}p} - {\textstyle \frac{1}{4}}\sum_{{\rm O}q} 
    I_{{\rm O}q\,{\rm D}p} \,\Bigr)\\ 
  \nonumber
  =& \hspace{6pt}
  \sum_{{\rm D}q} N_{{\rm D}q}
  \Bigl( I_{{\rm D}q\,{\rm D}p} + 
  I_{{\rm D}q'\,{\rm D}p} \Bigr) 
  + \sum_{{\rm O}q}   I_{{\rm O}q\,{\rm D}p} \\
  \label{anom_gen_1}
  =& \:\frac{1}{N_{{\rm D}p}} \int_{\mathcal{X}} 
  \biggl( \;\sum_{{\rm D}q,{\rm D}q'} 
  \mathcal{Q}\bigl(\,\Gamma_{{\rm D}q}\,, \ov\F^+_{{\rm D}q} \,\bigr)
  + \sum_{{\rm O}q}  
  \mathcal{Q}\bigl(\,\Gamma_{{\rm O}q}\, \bigr) \biggr)
  \wedge
  \mathcal{Q}\bigl(\,\Gamma_{{\rm D}p}\,, -\ov\F^+_{{\rm D}p} \,\bigr) \;.
\end{align}
Employing then the tadpole cancellation conditions \eqref{tadpole_compact_1} and \eqref{tadpole_compact_2} together with the explicit form of the charges \eqref{top_charges}, we see that the anomaly \eqref{anom_gen_1} can be simplified to
\eq{
  \mathcal{A}_{SU(N_{{\rm D}3})^3} &=
  \mathcal{A}_{SU(N_{{\rm D}5})^3} =
  \mathcal{A}_{SU(N_{{\rm D}7})^3} = 0\;, \\[3mm]
  \mathcal{A}_{SU(N_{{\rm D}9})^3} &=
  - \frac{\kappa_9}{N_{{\rm D}9}} \int_{\Gamma_{{\rm D}9}} 
  H_3\wedge F_3
  \;\overset{\rm Freed-Witten}{=}\; 0 \;.
}
For D$9$-branes, the cubic non-abelian anomaly vanishes due to the Free-Witten anomaly cancellation condition \cite{Freed:1999vc} which means that $H_3$ restricted to a D-brane has to be zero.

\bigskip
Along the same lines as in section \ref{sec_chiral_anom}, we can determine the mixed abelian--non-abelian, the cubic abelian and the mixed abelian--gravitational anomalies to be of the following form
\eq{
  \arraycolsep2pt
  \renewcommand{\arraystretch}{2}
  \begin{array}{lcrcr}
  \displaystyle \mathcal{A}_{U(1)_{{\rm D}p}-SU(N_{{\rm D}q})^2}
    &=& 
    \displaystyle \frac{1}{2} \:\delta_{{\rm D}p,{\rm D}q}\:
    \mathcal{A}_{SU(N_{{\rm D}p})^3}
    &-& \displaystyle  \frac{1}{2}\: N_{{\rm D}q} \:\Bigl( 
    I_{{\rm D}p\,{\rm D}q} - I_{{\rm D}p'{\rm D}q} \Bigr) \;,
  \\
  \displaystyle \mathcal{A}_{U(1)_{{\rm D}p}-U(1)^2_{{\rm D}q}}
    &=& 
    \displaystyle \frac{N_{{\rm D}p}}{3} \:\delta_{{\rm D}p,{\rm D}q}\:
    \mathcal{A}_{SU(N_{{\rm D}p})^3}
    & -& \displaystyle  N_{{\rm D}p}\:N_{{\rm D}q}\: \Bigl( 
    I_{{\rm D}p\,{\rm D}q} - I_{{\rm D}p'{\rm D}q} \Bigr) \;,
  \\
  \displaystyle \mathcal{A}_{U(1)_{{\rm D}p}-G^2}
    &=& \displaystyle  N_{{\rm D}p} \:
    \mathcal{A}_{SU(N_{{\rm D}p})^3}
    &-& \displaystyle  \frac{3}{4}\: N_{{\rm D}p}\: \sum_{{\rm O}q} 
    I_{{\rm O}q\,{\rm D}p} \;.\hspace{37pt}
  \end{array}
}
We are not going to show that the dimensional reduction of the Chern-Simons actions \eqref{action_cs}  provides the required Green--Schwarz couplings to cancel these anomalies. This can be done in a very similar way as in section \ref{sec_gsm}.

\vskip10mm

\subsection{Massive U(1)s and Fayet-Iliopoulos Terms}

We finish this section with a discussion of  massive $U(1)$ fields and the Fayet-Iliopoulos terms.
For the case of D$7$- and D$3$-branes, this has been done in section \ref{sec_mass_fi} so here we will focus on the D$5$- and D$9$-branes.
Furthermore, we will consider only diagonally embedded abelian fluxes on the D$5$- and D$9$-branes in order to simplify the discussion.

\bigskip
To determine the couplings of the $U(1)$ gauge bosons on the D$5$-branes to the R-R $p$-form potentials $C_p$, let us expand the two-cycle the D$5$-brane is wrapping as 
\eq{
  \Gamma_{{\rm D}5}= m_{{\rm D}5\,I}\:\Sigma^I \;,
}
where $\{\Sigma^I\}$ denotes the basis of two-cycles introduced in equation \eqref{basis_4_cycles}. 
Writing then out the Chern characters as in equation \eqref{chern_exp_1}, we obtain
\begin{align}
  \nonumber
  \mathcal{S}_{\rm mass} &= -\frac{\kappa_5}{l_s^2} \int_{\mathbb{R}^{3,1}}
  \sum_{a,a'} f_{{\rm D}5_a} \wedge 
  \biggl( N_{{\rm D}5_a} D^I\wedge \frac{1}{l_s^2} 
  \int_{\Gamma_{{\rm D}5_a}} \omega_I
  +
  \D_0 \wedge \frac{1}{l_s^2} \int_{\Gamma_{{\rm D}5_a}} 
  \ch_1\bigl( \ov \F^+_{{\rm D}5_a} \bigr) \biggr) \\
  &= -\frac{\kappa_5}{l_s^2} \int_{\mathbb{R}^{3,1}}
  \sum_{a} f_{{\rm D}5_a} \wedge 
  \biggl( N_{{\rm D}5_a} \bigl( m_{{\rm D}5_a\,I}
  + m_{{\rm D}5_{a'}\,I} \bigr) \,  D^I \biggr) \;,
\end{align}
where the  term involving $\ch_1(\ov\F^+)$ vanishes due to its orientifold image.
The mass matrix for the $U(1)$ gauge bosons on the D$5$-branes therefore reads
\eq{
  \label{mass_d5}
  M_{I_+\, {\rm D}5_a} = 
  N_{{\rm D}5_a} \bigl( m_{{\rm D}5_a}
  + m_{{\rm D}5_{a'}} \bigr)_{I_+} \;.
}
Finally, recalling the form of the derivative of the K\"ahler potential with respect to $T_{I_+}$ given in equation \eqref{kaehler_deriv} and noting that \eqref{mass_d5} corresponds to the holomorphic Killing vectors of the gauge isometry associated to $T_{I_+}$, we can determine the Fayet-Iliopoulos term of a D$5$-brane as
\eq{
   \xi_{{\rm D}5_a} \:\sim\: -i\,M_{I_+ {\rm D}5_a} \: 
    \frac{\partial \mathcal{K}}{\partial T_{I_+}}
   \:\sim\: \frac{1}{l_s^2}\:\frac{e^{\phi}}{\mathcal{V}}\: N_{{\rm D}5_a}
   \int_{\Gamma_{{\rm D}5_a}} J
   \;.
}

\bigskip
In order to study the mass matrix and Fayet-Iliopoulos terms for the D$9$-branes, let us expand the R-R eight-form potential $C_8$ in the following way
\eq{
  C_8 = D^0\wedge d{\rm vol}_{\mathcal{X}} \;.
}
Writing out the fourth Chern character similarly as in \eqref{chern_exp_1}, we can determine the  mass terms for the $U(1)$ gauge bosons on the D$9$-branes as
\eq{
  \label{mass_d9}
  \mathcal{S}_{\rm mass} =& -\frac{\kappa_9}{l_s^2} \int_{\mathbb{R}^{3,1}}
  \sum_{a,a'} f_{{\rm D}9_a} \wedge 
  \Biggl[ \quad 
  D^0\wedge \bigl[ \Gamma_{{\rm D}9_a} \bigr] \: N_{{\rm D}9_a} \\
  &\hspace{40pt}+ \D_I\wedge \frac{1}{l_s^6}\int_{\Gamma_{{\rm D}9_a}} 
    \sigma^I\wedge 
    \ch_1\bigl( \ov \F^+_{{\rm D}9_a} \bigr) \\
  &\hspace{40pt}+ D^I \wedge \frac{1}{l_s^6}\int_{\Gamma_{{\rm D}9_a}} 
    \omega_I\wedge  
    \left(
    \ch_2\bigl( \ov \F^+_{{\rm D}9_a} \bigr)
    + l_s^4\, N_{{\rm D}9_a}\: \frac{c_2\bigl( \mathcal{X} \bigr)}{24}
    \right)    \\
  &\hspace{40pt}+ \D_0 \wedge \frac{1}{l_s^6}\int_{\Gamma_{{\rm D}9_a}} 
    \left(
    \ch_3\bigl( \ov \F^+_{{\rm D}9_a} \bigr)
    + l_s^4\, \ch_1\bigl( \ov \F^+_{{\rm D}9_a} \bigr)
    \: \frac{c_2\bigl( \mathcal{X} \bigr)}{24}
    \right)    
  \quad \Biggr] .
}
Taking into account the explicit expression for the orientifold images, we see that the couplings $f_{{\rm D}9}\wedge \D_0$ in the last line of \eqref{mass_d9} vanish. From the remaining terms, we determine the following mass matrices
\begin{align}
  \nonumber
  f_{{\rm D}9_a} - D^0  :\quad 
  &M_{{\rm D}9_a} \hspace{10pt}=  2\: N_{{\rm D}9_a} \;, \\[4.5mm]
  f_{{\rm D}9_a} - \D_I :\quad 
  &M^{I_-}_{{\rm D}9_a} \hspace{10pt}=
  \frac{1}{l_s^6} \int_{\mathcal{X}} 
  \ch_1\bigl( \ov \F^+_{{\rm D}9_a}
  \bigr) \wedge \bigl( \sigma^I-\sigma^*\sigma^I \bigr)\;, \\[2mm]
  \nonumber
  f_{{\rm D}9_a} - D^I :\quad 
  &M_{I_+{\rm D}9_a} =
  \frac{1}{l_s^6} \int_{\mathcal{X}} 
    \left(
    \ch_2\bigl( \ov \F^+_{{\rm D}9_a} \bigr)
    + l_s^4\, N_{{\rm D}9_a}\: \frac{c_2\bigl( \mathcal{X} \bigr)}{24}
    \right)  \wedge \Bigl( \omega + \sigma^* \omega \Bigr)_{I_+}
    \;,
\end{align}
where $\sigma^*\sigma^I$ denotes the image of the basis four-form $\sigma^I$ under the holomorphic involution $\sigma$.
The Fayet-Iliopoulos terms for the D$9$-branes are then computed similarly as in the previous cases using the derivatives \eqref{kaehler_deriv} of the K\"ahler potential. Concretely, by employing \eqref{ch_2_exp} we find
\eq{
  \xi_{{\rm D}9_a} \:&\sim\: 
    -i\,M_{I_+{\rm D}9_a} \: \frac{\partial \mathcal{K}}{\partial T_{I_+}}
    -i\,M_{{\rm D}9_a}^{I_-} \: \frac{\partial \mathcal{K}}
    {\partial G^{I_-}} 
    +i\,M_{{\rm D}9_a} \: \frac{\partial \mathcal{K}}
    {\partial \tau} 
    \\
  &\sim\:
  \frac{e^{\phi}}{\mathcal{V}}\,
  \left[\: \frac{1}{l_s^6} \int_{\mathcal{X}}
   \biggl( \ch_2\bigl( \ov \F_{{\rm D}9_a} \bigr)
    + l_s^4\, N_{{\rm D}9_a}\: \frac{c_2\bigl( \mathcal{X} \bigr)}{24}
    \biggr)  \wedge J
    \;\;-\;\;\mathcal{V}\: \right] \;.
}


\section{Summary and Conclusions}
\label{sec_summary}

In this work, we have studied type IIB string theory compactifications on orientifolds of {\em smooth} compact Calabi-Yau manifolds with D$3$- and D$7$-branes. In particular, we have derived the tadpole cancellation conditions in detail and we have shown how the generalized Green--Schwarz mechanism cancels the chiral anomalies.
Of course, in accordance with results obtained for toroidal orbifolds, this was expected from the very beginning, however, the detailed study has lead to the following observations. 
\begin{itemize}

\item For an orientifold projection $\Omega(-1)^{F_L}\sigma$ leading to $h^{1,1}_-\neq0$, that is there are two- and four-cycles  anti-invariant under the holomorphic involution $\sigma$, in general the D$5$-brane tadpole cancellation condition leads to a non-trivial constraint. This has already been mentioned in \cite{Collinucci:2008pf}, however, here we have worked out this condition in detail.

\item We have furthermore seen that for the cancellation of chiral anomalies not only the D$7$-brane tadpole cancellation condition has to be employed, but in general also the vanishing of the induced D$5$-brane charges.

\item In section \ref{sec_generalization}, we have generalized our  analysis by including also D$9$- and D$5$-branes for which we have worked out the tadpole cancellation conditions in detail. Utilizing the requirement that the latter ensure the vanishing of the cubic non-abelian anomaly, we were able to determine a general set of rules for computing the chiral spectrum of the combined system of D$9$-, D$7$-, D$5$- and D$3$-branes. These have been summarized in table \ref{rule_spec_gen}.

\end{itemize}

The work presented in this paper is intended to provide a piece for a better understanding of the open sector of type IIB orientifold compactifications on smooth Calabi-Yau manifolds with D$3$- and D$7$-branes. In particular, we have seen that not only the well-known D$3$- and D$7$-brane tadpole cancellation conditions arise in such setups, but that in general also the cancellation of the induced D$5$-brane charge is crucial for the consistency of a (compact) model. Clearly, this observation has to be taken into account when embedding local F-theory models into compact Calabi-Yau manifolds.

We have also observed that including D$9$- and D$5$-branes leads to a more involved structure of the open sector. 
However, it might be interesting to study the combined system of D$9$-, D$7$-, D$5$- and D$3$-branes in type IIB orientifolds with O$7$- and O$3$-planes in more detail, and to work out its relation to F-theory. This could lead to a better understanding of the connection between type IIB string theory and F-theory.


\vskip15mm

\subsubsection*{Acknowledgments}

The author would like to thank Nikolas Akerblom, Sebastian Moster, Maximilian Schmidt-Sommerfeld and especially Ralph Blumenhagen, Thomas Grimm and Timo Weigand for  helpful discussions. Furthermore, he wants to thank Maximilian Schmidt-Sommerfeld for useful comments on the manuscript and Dieter L\"ust for support. 

\medskip
In addition, the author would like to thank the referee at JHEP for drawing his attention to earlier work on the generalized Green--Schwarz mechanism and tadpole cancellation conditions in the context of toroidal type IIB orientifolds.


\clearpage
\begin{appendix}


\section{More Details on the Chern-Simons Action}
\label{app_cs_action}

In this appendix, we provide the definitions of the quantities used in the Chern-Simons actions \eqref{action_cs} for D-branes and O-planes, and give some details of the calculation leading to \eqref{top_exp_1}.


\subsubsection*{Definitions}

We start with the definitions.
The Chern character of a complex vector bundle $F$ is defined in the following way
\eq{
  \label{def_ch}
  \ch\bigl( F \bigr) = \sum_{n=0}^{\infty} \ch_n\bigl( F\bigr) 
  \;, \hspace{40pt}
  \ch_n\bigl( F\bigr) = \frac{1}{n!}\:\mbox{tr} \left[
  \left( \frac{iF}{2\pi}  \right)^n\right] \;,
}
where the trace is over the fundamental representation. The Chern character  satisfies 
\begin{align}
  \label{chern_ch_01}  
  \ch\,\bigl( E\oplus F \bigr) & = \ch\bigl(E\bigr)+\ch\bigl( F\bigr) 
  \;.
\end{align}
The $\A$-genus and the Hirzebruch $\mathcal{L}$-polynomial can be expressed in terms of the Pontrjagin classes $p_i$ as 
\eq{
  \arraycolsep1.5pt
  \begin{array}{lcrrcrll}
  \displaystyle \A\bigl( F \bigr) & 
    = & 
    1 - &
    \displaystyle \frac{1}{24}\,p_1 +& 
    \displaystyle \frac{1}{5760}&
    \displaystyle \bigl( 7 p_1^2-4 p_2 \bigr) &
    + \ldots \;,\\[3mm]
  \displaystyle \mathcal{L}\bigl( F \bigr) &
    = &
    1 + &
    \displaystyle \frac{1}{3}\, p_1 +& 
    \displaystyle \frac{1}{45}&
    \displaystyle \bigl( - p_1^2+7 p_2 \bigr) &
    + \ldots \;,
  \end{array}
}
and satisfy
\eq{
  \label{app_splitting_2}
  \A\bigl( E \oplus F \bigr) = \A\bigl( E \bigr) \wedge \A\bigl( F \bigr)
  \;,\hspace{40pt}
  \mathcal{L}\bigl( E \oplus F \bigr) = \mathcal{L}\bigl( E \bigr)
  \wedge \mathcal{L}\bigl( F \bigr) \;.
}
For the following, we will only need the definition of the first Pontrjagin class of a real vector bundle which reads
\eq{
  \label{def_p1}
  p_1\bigl( F \bigr) =
  -\,\frac{1}{2}\: \mbox{tr} \left[ \left(
  \frac{F}{2\pi} \right)^2 \right] \;,
}
where the trace is again over the fundamental representation.
If the real $2k$-dimensional bundle $F_{\mathbb{R}}$ can be written as a complex $k$-dimensional bundle $F_{\mathbb{C}}$, we have the relation
\eq{
  \label{app_p1_c}
  p_1\bigl( F_{\mathbb{R}} \bigr)= \left[c_1\bigl( F_{\mathbb{C}}\bigr)
  \right]^2- 
  2\,c_2\bigl(F_{\mathbb{C}}\bigr)\;,
}  
where $c_1$ and $c_2$ denote the first and second Chern class expressed as
\eq{
  c_1\bigl( F \bigr) = \ch_1\bigl( F \bigr)
  \;,\hspace{40pt}
  c_2\bigl( F \bigr) = \frac{1}{2}\:\Bigl[ 
    \ch_1\bigl( F \bigr) \Bigr]^2
    - \ch_2\bigl( F \bigr) \;.
}


\subsubsection*{Calculation leading to (\ref{top_exp_1})}

After stating these definitions and relations, let us concentrate on a complex two-dimensional holomorphic submanifold $\Gamma$ of a complex three-dimensional Calabi-Yau manifold $\mathcal{X}$. Since the first Chern class of a Calabi-Yau manifold vanishes, we find
\eq{
  \label{app_relation1}
  0 = c_1\bigl( T_{\mathcal{X}} \bigr) 
  = \ch_1\bigl( T_{\Gamma}\oplus N_{\Gamma} \bigr) 
  = \ch_1\bigl( T_{\Gamma}\bigr) + \ch_1\bigl( N_{\Gamma} \bigr)
  = c_1 \bigl( T_{\Gamma}\bigr) + c_1\bigl( N_{\Gamma} \bigr) \;,
}
where $T$ denotes the tangential bundle and $N$ the normal bundle.
Noting then that the second Chern class of a line bundle such as $N_{\Gamma}$ vanishes, we calculate using \eqref{app_p1_c} and \eqref{app_relation1}
\eq{
  p_1\bigl( T_{\Gamma} \bigr) - p_1\bigl( N_{\Gamma} \bigr)
  = \Bigl[c_1\bigl(T_{\Gamma}\bigr)\Bigr]^2- 2\,c_2\bigl( T_{\Gamma}\bigr)
    -\Bigl[c_1\bigl(N_{\Gamma}\bigr)\Bigr]^2+ 2\,c_2\bigl( N_{\Gamma}\bigr) 
  = - 2\,c_2\bigl( T_{\Gamma} \bigr)
}
where we interpreted the real vector bundles as complex ones.
This computation allows us now to write the $\A$-terms in the Chern-Simons action more feasible. The square root as well as the inverse of the $\A$-genus are understood as a series expansion and using \eqref{app_splitting_2}, we find
\eq{
  &\sqrt{\frac{\A(\R_T)}{\A (\R_N)}} 
  = \sqrt{\A(\R^{(4)})}\wedge \sqrt{\frac{\A(\R^{(6)}_T)}{\A (\R^{(6)}_N)}}
  \\
  =& \left( 1 - \frac{1}{48}\, p_1\bigl( \R^{(4)}\bigr)+\ldots\right)
    \wedge 
    \left( 1 - \frac{1}{48}\, p_1\bigl( \R_T^{(6)}\bigr)
    + \frac{1}{48}\, p_1\bigl( \R_N^{(6)}\bigr) +\ldots\right) \\
  =& \left( 1 + \frac{1}{96} \left(\frac{l_s^2}{2\pi}\right)^2
    \!\mbox{tr}\,\bigl( R^2\bigr)
    +\ldots\right)
    \wedge 
    \Biggl( 1 + \frac{l_s^4}{24}\: c_2\bigl( \Gamma\bigr) + 
    \ldots\Biggr) \;,
}
where \textsuperscript{(4)} denotes the four-dimensional and \textsuperscript{(6)} the internal part of $\R$. In going from the second to the third line, we employed our definition \eqref{def_r} and we adjusted our notation as
\eq{
  \arraycolsep2pt
  \begin{array}{lclclr}
  \displaystyle p_1\bigl(\R_T^{(6)}\bigr) &=&
    \displaystyle l_s^4\:p_1\bigl(\ov R_T\bigr)&=&
      \displaystyle l_s^4\:p_1\bigl( T_{\Gamma}\bigr)\;, 
  \hspace{40pt}\displaystyle c_2\bigl( T_{\Gamma} \bigr) 
  = c_2\bigl( \Gamma \bigr) \;, \\[2mm]
  \displaystyle p_1\bigl(\R_N^{(6)}\bigr)&=&
    \displaystyle l_s^4\:p_1\bigl(\ov R_N\bigr)&=&
      \displaystyle l_s^4\:p_1\bigl( N_{\Gamma}\bigr) \;.
  \end{array}
}  
Along the same lines, we obtain for the Hirzebruch $\mathcal{L}$-polynomial the following result
\eq{
  \sqrt{\frac{\mathcal{L}(\R_T/4)}{\mathcal{L} (\R_N/4)}} 
  = \left( 1 - \frac{1}{192} \left(\frac{l_s^2}{2\pi}\right)^2
    \!\mbox{tr}\,\bigl( R^2\bigr)
    +\ldots\right)
    \wedge 
    \Biggl( 1 - \frac{l_s^4}{48}\: c_2\bigl( \Gamma\bigr) + 
    \ldots\Biggr) .
}


\section{Discussion for SO(2N) and Sp(2N)}
\label{app_sp_so}

Here, we briefly discuss the generalized Green--Schwarz mechanism for the case of gauge groups $SO(2N)$ and $Sp(2N)$. Since both Lie groups are simple, there are no cubic abelian or mixed abelian--gravitational anomalies for these cases. 
For the cubic non-abelian anomaly, let us note that the anomaly is proportional to 
\eq{
  \mathcal{A}^{abc}(r) = \frac{1}{2}\:A(r)\:d^{abc}
}
where $d^{abc}$ is the unique symmetric invariant. This invariant only exists for $SU(N)$ and $SO(6)$ (which has the same Lie algebra as $SU(4)$) and so there is no cubic non-abelian anomaly to be studied in the present case.

For the mixed abelian--non-abelian anomaly, let us note that the dimension and the index for the fundamental representation of both $SO(2N)$ and $Sp(2N)$ are found to be
\eq{
  \mbox{dim}\bigl( F \bigr)= 2N\;, 
  \hspace{60pt}
  C\bigl( F \bigr) = 1 \;.
}
The anomaly coefficient is then computed as 
\eq{
  \label{app_anom_mana}
   \mathcal{A}_{U(1)_a-Sp/SO(2N_{{\rm D7}_b})^2} &= 
   \sum_F Q_a\bigl( F \bigr)\: C_b\bigl( F \bigr)
   =  -N_{{\rm D7}_a}\, \Bigl( I_{ab} - I_{a'b} \Bigr) \;,
}
which is, up to a factor of $\frac{1}{2}$, the same as in \eqref{anom_mana}. For the calculation of the Green-Schwarz diagrams, we note that $C(F)=1$ by definition means
\eq{
  \mbox{tr}\,\bigl( T^A\,T^B \bigr)= \delta^{AB} \;,
}
which differs from the result for $SU(N)$ by a factor of $\frac{1}{2}$. Using this observation and following the same steps as in the computation for $SU(N)$, one finds that the Green--Schwarz diagrams are precisely of the form \eqref{app_anom_mana} (up to a common prefactor). Therefore, also for $SO(2N)$ and $Sp(2N)$ the mixed abelian--non-abelian anomalies are cancelled via the generalized Green-Schwarz mechanism.

\end{appendix}


\clearpage
\nocite{*}
\bibliography{references}

\providecommand{\href}[2]{#2}\begingroup\raggedright\begin{thebibliography}{10}

\bibitem{Blumenhagen:2005mu}
R.~Blumenhagen, M.~Cveti\v{c}, P.~Langacker, and G.~Shiu, ``{Toward realistic
  intersecting D-brane models},'' {\em Ann. Rev. Nucl. Part. Sci.} {\bf 55}
  (2005) 71--139,
\href{http://www.arXiv.org/abs/hep-th/0502005}{{\tt hep-th/0502005}}.

\bibitem{Green:1984sg}
M.~B. Green and J.~H. Schwarz, ``{Anomaly Cancellation in Supersymmetric D=10
  Gauge Theory and Superstring Theory},'' {\em Phys. Lett.} {\bf B149} (1984)
117--122.

\bibitem{Sagnotti:1992qw}
A.~Sagnotti, ``{A Note on the Green-Schwarz mechanism in open string
  theories},'' {\em Phys. Lett.} {\bf B294} (1992) 196--203,
\href{http://www.arXiv.org/abs/hep-th/9210127}{{\tt hep-th/9210127}}.

\bibitem{Aldazabal:1998mr}
G.~Aldazabal, A.~Font, L.~E. Ibanez, and G.~Violero, ``{D = 4, N = 1, type IIB
  orientifolds},'' {\em Nucl. Phys.} {\bf B536} (1998) 29--68,
\href{http://www.arXiv.org/abs/hep-th/9804026}{{\tt hep-th/9804026}}.

\bibitem{Ibanez:1998qp}
L.~E. Ibanez, R.~Rabadan, and A.~M. Uranga, ``{Anomalous U(1)'s in type I and
  type IIB D = 4, N = 1 string vacua},'' {\em Nucl. Phys.} {\bf B542} (1999)
  112--138,
\href{http://www.arXiv.org/abs/hep-th/9808139}{{\tt hep-th/9808139}}.

\bibitem{Aldazabal:1999nu}
G.~Aldazabal, D.~Badagnani, L.~E. Ibanez, and A.~M. Uranga, ``{Tadpole versus
  anomaly cancellation in D = 4, 6 compact IIB orientifolds},'' {\em JHEP} {\bf
  06} (1999) 031,
\href{http://www.arXiv.org/abs/hep-th/9904071}{{\tt hep-th/9904071}}.

\bibitem{Ibanez:1999pw}
L.~E. Ibanez, R.~Rabadan, and A.~M. Uranga, ``{Sigma-model anomalies in compact
  D = 4, N = 1 type IIB orientifolds and Fayet-Iliopoulos terms},'' {\em Nucl.
  Phys.} {\bf B576} (2000) 285--312,
\href{http://www.arXiv.org/abs/hep-th/9905098}{{\tt hep-th/9905098}}.

\bibitem{Scrucca:1999zh}
C.~A. Scrucca and M.~Serone, ``{Gauge and gravitational anomalies in D = 4 N =
  1 orientifolds},'' {\em JHEP} {\bf 12} (1999) 024,
\href{http://www.arXiv.org/abs/hep-th/9912108}{{\tt hep-th/9912108}}.

\bibitem{Aldazabal:2000dg}
G.~Aldazabal, S.~Franco, L.~E. Ibanez, R.~Rabadan, and A.~M. Uranga, ``{D = 4
  chiral string compactifications from intersecting branes},'' {\em J. Math.
  Phys.} {\bf 42} (2001) 3103--3126,
\href{http://www.arXiv.org/abs/hep-th/0011073}{{\tt hep-th/0011073}}.

\bibitem{Witten:1982fp}
E.~Witten, ``{An SU(2) anomaly},'' {\em Phys. Lett.} {\bf B117} (1982)
324--328.

\bibitem{Uranga:2000xp}
A.~M. Uranga, ``{D-brane probes, RR tadpole cancellation and K-theory
  charge},'' {\em Nucl. Phys.} {\bf B598} (2001) 225--246,
\href{http://www.arXiv.org/abs/hep-th/0011048}{{\tt hep-th/0011048}}.

\bibitem{Blumenhagen:2005zh}
R.~Blumenhagen, G.~Honecker, and T.~Weigand, ``{Non-abelian brane worlds: The
  open string story},''
\href{http://www.arXiv.org/abs/hep-th/0510050}{{\tt hep-th/0510050}}.

\bibitem{Bachas:2008jv}
C.~Bachas, M.~Bianchi, R.~Blumenhagen, D.~L{\"u}st, and T.~Weigand, ``{Comments
  on Orientifolds without Vector Structure},'' {\em JHEP} {\bf 08} (2008) 016,
\href{http://www.arXiv.org/abs/0805.3696}{{\tt 0805.3696}}.

\bibitem{Jockers:2004yj}
H.~Jockers and J.~Louis, ``{The effective action of D7-branes in N = 1
  Calabi-Yau orientifolds},'' {\em Nucl. Phys.} {\bf B705} (2005) 167--211,
\href{http://www.arXiv.org/abs/hep-th/0409098}{{\tt hep-th/0409098}}.

\bibitem{Jockers:2005zy}
H.~Jockers and J.~Louis, ``{D-terms and F-terms from D7-brane fluxes},'' {\em
  Nucl. Phys.} {\bf B718} (2005) 203--246,
\href{http://www.arXiv.org/abs/hep-th/0502059}{{\tt hep-th/0502059}}.

\bibitem{Collinucci:2008pf}
A.~Collinucci, F.~Denef, and M.~Esole, ``{D-brane Deconstructions in IIB
  Orientifolds},''
\href{http://www.arXiv.org/abs/0805.1573}{{\tt 0805.1573}}.

\bibitem{Pradisi:1988xd}
G.~Pradisi and A.~Sagnotti, ``{Open String Orbifolds},'' {\em Phys. Lett.} {\bf
  B216} (1989)
59.

\bibitem{Gimon:1996rq}
E.~G. Gimon and J.~Polchinski, ``{Consistency Conditions for Orientifolds and
  D-Manifolds},'' {\em Phys. Rev.} {\bf D54} (1996) 1667--1676,
\href{http://www.arXiv.org/abs/hep-th/9601038}{{\tt hep-th/9601038}}.

\bibitem{Angelantonj:1996uy}
C.~Angelantonj, M.~Bianchi, G.~Pradisi, A.~Sagnotti, and Y.~S. Stanev,
  ``{Chiral asymmetry in four-dimensional open- string vacua},'' {\em Phys.
  Lett.} {\bf B385} (1996) 96--102,
\href{http://www.arXiv.org/abs/hep-th/9606169}{{\tt hep-th/9606169}}.

\bibitem{Kakushadze:1997ku}
Z.~Kakushadze and G.~Shiu, ``{A chiral N = 1 type I vacuum in four dimensions
  and its heterotic dual},'' {\em Phys. Rev.} {\bf D56} (1997) 3686--3697,
\href{http://www.arXiv.org/abs/hep-th/9705163}{{\tt hep-th/9705163}}.

\bibitem{Bianchi:2000de}
M.~Bianchi and J.~F. Morales, ``{Anomalies and tadpoles},'' {\em JHEP} {\bf 03}
  (2000) 030,
\href{http://www.arXiv.org/abs/hep-th/0002149}{{\tt hep-th/0002149}}.

\bibitem{Kachru:2003aw}
S.~Kachru, R.~Kallosh, A.~Linde, and S.~P. Trivedi, ``{De Sitter vacua in
  string theory},'' {\em Phys. Rev.} {\bf D68} (2003) 046005,
\href{http://www.arXiv.org/abs/hep-th/0301240}{{\tt hep-th/0301240}}.

\bibitem{Balasubramanian:2005zx}
V.~Balasubramanian, P.~Berglund, J.~P. Conlon, and F.~Quevedo, ``{Systematics
  of moduli stabilisation in Calabi-Yau flux compactifications},'' {\em JHEP}
  {\bf 03} (2005) 007,
\href{http://www.arXiv.org/abs/hep-th/0502058}{{\tt hep-th/0502058}}.

\bibitem{Conlon:2005ki}
J.~P. Conlon, F.~Quevedo, and K.~Suruliz, ``{Large-volume flux
  compactifications: Moduli spectrum and D3/D7 soft supersymmetry breaking},''
  {\em JHEP} {\bf 08} (2005) 007,
\href{http://www.arXiv.org/abs/hep-th/0505076}{{\tt hep-th/0505076}}.

\bibitem{Conlon:2008wa}
J.~P. Conlon, A.~Maharana, and F.~Quevedo, ``{Towards Realistic String
  Vacua},''
\href{http://www.arXiv.org/abs/0810.5660}{{\tt 0810.5660}}.

\bibitem{Blumenhagen:2007sm}
R.~Blumenhagen, S.~Moster, and E.~Plauschinn, ``{Moduli Stabilisation versus
  Chirality for MSSM like Type IIB Orientifolds},'' {\em JHEP} {\bf 01} (2008)
  058,
\href{http://www.arXiv.org/abs/0711.3389}{{\tt 0711.3389}}.

\bibitem{Vafa:1996xn}
C.~Vafa, ``{Evidence for F-Theory},'' {\em Nucl. Phys.} {\bf B469} (1996)
  403--418,
\href{http://www.arXiv.org/abs/hep-th/9602022}{{\tt hep-th/9602022}}.

\bibitem{Beasley:2008dc}
C.~Beasley, J.~J. Heckman, and C.~Vafa, ``{GUTs and Exceptional Branes in
  F-theory - I},''
\href{http://www.arXiv.org/abs/0802.3391}{{\tt 0802.3391}}.

\bibitem{Buchbinder:2008at}
E.~I. Buchbinder, ``{Dynamically SUSY Breaking SQCD on F-Theory
  Seven-Branes},''
\href{http://www.arXiv.org/abs/0805.3157}{{\tt 0805.3157}}.

\bibitem{Beasley:2008kw}
C.~Beasley, J.~J. Heckman, and C.~Vafa, ``{GUTs and Exceptional Branes in
  F-theory - II: Experimental Predictions},''
\href{http://www.arXiv.org/abs/0806.0102}{{\tt 0806.0102}}.

\bibitem{Heckman:2008es}
J.~J. Heckman, J.~Marsano, N.~Saulina, S.~Schafer-Nameki, and C.~Vafa,
  ``{Instantons and SUSY breaking in F-theory},''
\href{http://www.arXiv.org/abs/0808.1286}{{\tt 0808.1286}}.

\bibitem{Marsano:2008jq}
J.~Marsano, N.~Saulina, and S.~Schafer-Nameki, ``{Gauge Mediation in F-Theory
  GUT Models},''
\href{http://www.arXiv.org/abs/0808.1571}{{\tt 0808.1571}}.

\bibitem{Donagi:2008kj}
R.~Donagi and M.~Wijnholt, ``{Breaking GUT Groups in F-Theory},''
\href{http://www.arXiv.org/abs/0808.2223}{{\tt 0808.2223}}.

\bibitem{Marsano:2008py}
J.~Marsano, N.~Saulina, and S.~Schafer-Nameki, ``{An Instanton Toolbox for
  F-Theory Model Building},''
\href{http://www.arXiv.org/abs/0808.2450}{{\tt 0808.2450}}.

\bibitem{Heckman:2008qt}
J.~J. Heckman and C.~Vafa, ``{F-theory, GUTs, and the Weak Scale},''
\href{http://www.arXiv.org/abs/0809.1098}{{\tt 0809.1098}}.

\bibitem{Wijnholt:2008db}
M.~Wijnholt, ``{F-Theory, GUTs and Chiral Matter},''
\href{http://www.arXiv.org/abs/0809.3878}{{\tt 0809.3878}}.

\bibitem{Font:2008id}
A.~Font and L.~E. Ibanez, ``{Yukawa Structure from U(1) Fluxes in F-theory
  Grand Unification},''
\href{http://www.arXiv.org/abs/0811.2157}{{\tt 0811.2157}}.

\bibitem{Heckman:2008qa}
J.~J. Heckman and C.~Vafa, ``{Flavor Hierarchy From F-theory},''
\href{http://www.arXiv.org/abs/0811.2417}{{\tt 0811.2417}}.

\bibitem{Blumenhagen:2008zz}
R.~Blumenhagen, V.~Braun, T.~W. Grimm, and T.~Weigand, ``{GUTs in Type IIB
  Orientifold Compactifications},''
\href{http://www.arXiv.org/abs/0811.2936}{{\tt 0811.2936}}.

\bibitem{Braun:2008pz}
A.~P. Braun, A.~Hebecker, C.~Ludeling, and R.~Valandro, ``{Fixing D7 Brane
  Positions by F-Theory Fluxes},''
\href{http://www.arXiv.org/abs/0811.2416}{{\tt 0811.2416}}.

\bibitem{Acharya:2002ag}
B.~Acharya, M.~Aganagic, K.~Hori, and C.~Vafa, ``{Orientifolds, mirror symmetry
  and superpotentials},''
\href{http://www.arXiv.org/abs/hep-th/0202208}{{\tt hep-th/0202208}}.

\bibitem{Brunner:2003zm}
I.~Brunner and K.~Hori, ``{Orientifolds and mirror symmetry},'' {\em JHEP} {\bf
  11} (2004) 005,
\href{http://www.arXiv.org/abs/hep-th/0303135}{{\tt hep-th/0303135}}.

\bibitem{Grimm:2004uq}
T.~W. Grimm and J.~Louis, ``{The effective action of N = 1 Calabi-Yau
  orientifolds},'' {\em Nucl. Phys.} {\bf B699} (2004) 387--426,
\href{http://www.arXiv.org/abs/hep-th/0403067}{{\tt hep-th/0403067}}.

\bibitem{Bergshoeff:2001pv}
E.~Bergshoeff, R.~Kallosh, T.~Ortin, D.~Roest, and A.~Van~Proeyen, ``{New
  formulations of D = 10 supersymmetry and D8 - O8 domain walls},'' {\em Class.
  Quant. Grav.} {\bf 18} (2001) 3359--3382,
\href{http://www.arXiv.org/abs/hep-th/0103233}{{\tt hep-th/0103233}}.

\bibitem{Giddings:2001yu}
S.~B. Giddings, S.~Kachru, and J.~Polchinski, ``{Hierarchies from fluxes in
  string compactifications},'' {\em Phys. Rev.} {\bf D66} (2002) 106006,
\href{http://www.arXiv.org/abs/hep-th/0105097}{{\tt hep-th/0105097}}.

\bibitem{Kachru:2002he}
S.~Kachru, M.~B. Schulz, and S.~Trivedi, ``{Moduli stabilization from fluxes in
  a simple IIB orientifold},'' {\em JHEP} {\bf 10} (2003) 007,
\href{http://www.arXiv.org/abs/hep-th/0201028}{{\tt hep-th/0201028}}.

\bibitem{Frey:2002hf}
A.~R. Frey and J.~Polchinski, ``{N = 3 warped compactifications},'' {\em Phys.
  Rev.} {\bf D65} (2002) 126009,
\href{http://www.arXiv.org/abs/hep-th/0201029}{{\tt hep-th/0201029}}.

\bibitem{Dasgupta:1999ss}
K.~Dasgupta, G.~Rajesh, and S.~Sethi, ``{M theory, orientifolds and G-flux},''
  {\em JHEP} {\bf 08} (1999) 023,
\href{http://www.arXiv.org/abs/hep-th/9908088}{{\tt hep-th/9908088}}.

\bibitem{Marino:1999af}
M.~Marino, R.~Minasian, G.~W. Moore, and A.~Strominger, ``{Nonlinear instantons
  from supersymmetric p-branes},'' {\em JHEP} {\bf 01} (2000) 005,
\href{http://www.arXiv.org/abs/hep-th/9911206}{{\tt hep-th/9911206}}.

\bibitem{Hristov:2008if}
K.~Hristov, ``{Axion Stabilization in Type IIB Flux Compactifications},''
\href{http://www.arXiv.org/abs/0810.3329}{{\tt 0810.3329}}.

\bibitem{Douglas:1995bn}
M.~R. Douglas, ``{Branes within branes},''
\href{http://www.arXiv.org/abs/hep-th/9512077}{{\tt hep-th/9512077}}.

\bibitem{Green:1996dd}
M.~B. Green, J.~A. Harvey, and G.~W. Moore, ``{I-brane inflow and anomalous
  couplings on D-branes},'' {\em Class. Quant. Grav.} {\bf 14} (1997) 47--52,
\href{http://www.arXiv.org/abs/hep-th/9605033}{{\tt hep-th/9605033}}.

\bibitem{Cheung:1997az}
Y.-K.~E. Cheung and Z.~Yin, ``{Anomalies, branes, and currents},'' {\em Nucl.
  Phys.} {\bf B517} (1998) 69--91,
\href{http://www.arXiv.org/abs/hep-th/9710206}{{\tt hep-th/9710206}}.

\bibitem{Morales:1998ux}
J.~F. Morales, C.~A. Scrucca, and M.~Serone, ``{Anomalous couplings for
  D-branes and O-planes},'' {\em Nucl. Phys.} {\bf B552} (1999) 291--315,
\href{http://www.arXiv.org/abs/hep-th/9812071}{{\tt hep-th/9812071}}.

\bibitem{Stefanski:1998yx}
J.~Stefanski, Bogdan, ``{Gravitational couplings of D-branes and O-planes},''
  {\em Nucl. Phys.} {\bf B548} (1999) 275--290,
\href{http://www.arXiv.org/abs/hep-th/9812088}{{\tt hep-th/9812088}}.

\bibitem{Scrucca:1999uz}
C.~A. Scrucca and M.~Serone, ``{Anomalies and inflow on D-branes and
  O-planes},'' {\em Nucl. Phys.} {\bf B556} (1999) 197--221,
\href{http://www.arXiv.org/abs/hep-th/9903145}{{\tt hep-th/9903145}}.

\bibitem{Blumenhagen:2006ci}
R.~Blumenhagen, B.~K{\"o}rs, D.~L{\"u}st, and S.~Stieberger,
  ``{Four-dimensional String Compactifications with D-Branes, Orientifolds and
  Fluxes},'' {\em Phys. Rept.} {\bf 445} (2007) 1--193,
\href{http://www.arXiv.org/abs/hep-th/0610327}{{\tt hep-th/0610327}}.

\bibitem{Braun:2008ua}
A.~P. Braun, A.~Hebecker, and H.~Triendl, ``{D7-Brane Motion from M-Theory
  Cycles and Obstructions in the Weak Coupling Limit},'' {\em Nucl. Phys.} {\bf
  B800} (2008) 298--329,
\href{http://www.arXiv.org/abs/0801.2163}{{\tt 0801.2163}}.

\bibitem{Aluffi:2007sx}
P.~Aluffi and M.~Esole, ``{Chern class identities from tadpole matching in type
  IIB and F-theory},''
\href{http://www.arXiv.org/abs/0710.2544}{{\tt 0710.2544}}.

\bibitem{MarchesanoBuznego:2003hp}
F.~G. Marchesano~Buznego, ``{Intersecting D-brane models},''
\href{http://www.arXiv.org/abs/hep-th/0307252}{{\tt hep-th/0307252}}.

\bibitem{Minasian:1997mm}
R.~Minasian and G.~W. Moore, ``{K-theory and Ramond-Ramond charge},'' {\em
  JHEP} {\bf 11} (1997) 002,
\href{http://www.arXiv.org/abs/hep-th/9710230}{{\tt hep-th/9710230}}.

\bibitem{Katz:2002gh}
S.~H. Katz and E.~Sharpe, ``{D-branes, open string vertex operators, and Ext
  groups},'' {\em Adv. Theor. Math. Phys.} {\bf 6} (2003) 979--1030,
\href{http://www.arXiv.org/abs/hep-th/0208104}{{\tt hep-th/0208104}}.

\bibitem{Aspinwall:2004jr}
P.~S. Aspinwall, ``{D-branes on Calabi-Yau manifolds},''
\href{http://www.arXiv.org/abs/hep-th/0403166}{{\tt hep-th/0403166}}.

\bibitem{Marchesano:2007de}
F.~Marchesano, ``{Progress in D-brane model building},'' {\em Fortsch. Phys.}
  {\bf 55} (2007) 491--518,
\href{http://www.arXiv.org/abs/hep-th/0702094}{{\tt hep-th/0702094}}.

\bibitem{Grimm:2007hs}
T.~W. Grimm, ``{Axion Inflation in Type II String Theory},'' {\em Phys. Rev.}
  {\bf D77} (2008) 126007,
\href{http://www.arXiv.org/abs/0710.3883}{{\tt 0710.3883}}.

\bibitem{Freed:1999vc}
D.~S. Freed and E.~Witten, ``{Anomalies in string theory with D-branes},''
\href{http://www.arXiv.org/abs/hep-th/9907189}{{\tt hep-th/9907189}}.

\end{thebibliography}\endgroup
\bibliographystyle{utphys}


\end{document}